\documentclass[twocolumn,showpacs,nofootinbib,preprintnumbers,superscriptaddress]{revtex4-1}
\usepackage{array}
\usepackage{color}
\usepackage{dcolumn}
\usepackage{bm}
\usepackage{epsf}
\usepackage{graphicx}
\usepackage{amsmath}
\usepackage{verbatim}
\usepackage{slashed}
\Huge

\newcommand{\beq}{\begin{equation}}
\newcommand{\eeq}{\end{equation}}
\newcommand{\bean}{\begin{eqnarray*}}
\newcommand{\eean}{\end{eqnarray*}\noindent}
\newcommand{\bea}{\begin{eqnarray}}
\newcommand{\eea}{\end{eqnarray}\noindent}

\begin{document}
\topmargin0in
\textheight 8.in 
\bibliographystyle{apsrev}
\title{Extended evolution equations for neutrino propagation in astrophysical and cosmological environments}

\author{Cristina Volpe}
\email{volpe@apc.univ-paris7.fr}
\affiliation{AstroParticule et Cosmologie (APC), CNRS/Universit\'e de Paris 7, 10, rue Alice Domon et L\'eonie Duquet, 75205 Paris cedex 13, France}

\author{Daavid V\"a\"an\"anen}
\email{vaananen@ipno.in2p3.fr}
\affiliation{AstroParticule et Cosmologie (APC), CNRS/Universit\'e de Paris 7, 10, rue Alice Domon et L\'eonie Duquet, 75205 Paris cedex 13, France}

\author{Catalina Espinoza}
\email{m.catalina@cftp.ist.utl.pt}
\affiliation{Centro de F{\'{\i}}sica Te\'orica de Part{\'{\i}}culas,
Instituto Superior T\'ecnico, Universidade T\'ecnica de Lisboa,
Av. Rovisco Pais 1, 1049-001 Lisboa, Portugal.}

\affiliation{Departamento de F{\'{i}}sica Te\'orica and IFIC, Universidad de
Valencia-CSIC, E-46100, Valencia, Spain.}

\date{\today}


\begin{abstract}
We derive the evolution equations for a system of neutrinos interacting among themselves and with a matter background, based upon the Bogoliubov-Born-Green-Kirkwood-Yvon (BBGKY) hierarchy.
This theoretical framework gives an (unclosed) set of first-order coupled  integro-differential equations governing the evolution of the reduced density matrices.
By employing the hierarchy, 
we first rederive the mean-field evolution equations for the neutrino one-body density matrix associated with a system of neutrinos and antineutrinos
interacting with matter and with an anisotropic neutrino background. Then,
 we derive extended evolution equations to determine neutrino flavor conversion beyond the commonly used mean-field approximation.
To this aim we include neutrino-antineutrino pairing correlations to the two-body density matrix. The inclusion of these new contributions leads to an extended evolution equation for the normal neutrino density and to an equation 
for the abnormal one involving the pairing mean-field. We discuss the possible impact of neutrino-antineutrino correlations on neutrino flavor conversion in the astrophysical and cosmological environments, and possibly upon the supernova dynamics. 
Our results can be easily generalized to an arbitrary number of neutrino families. 
\end{abstract}

\maketitle


\section{Introduction} \label{sec:intro}
\noindent
Oscillations between quantum states is a widespread phenomenon, appearing in different physical contexts, like
the Rabi oscillations in optics, or the $K_0-\bar{K}_0$ and neutrino oscillations in particle physics. The propagation in a medium can produce resonant conversion between quantum states, such as when neutrinos change their flavor while traveling in a star, or  when photons modify their polarization in a birefringent medium. The basic oscillation phenomenon can be modified sometimes in surprising ways, if the interaction with a medium introduces complexity. The study of how neutrinos change their flavor in stellar environments and in the early universe has uncovered numerous such examples. Complexity arises from the non-linear and the many-body character of the problem. 

Neutrinos are elementary particles having non-zero mixings, as first conjectured by Pontecorvo \cite{Pontecorvo:1957cp} and discovered in 1998 by Super-Kamiokande \cite{Fukuda:1998ah}. 
Numerous experiments have contributed to
the measurement of the neutrino mixing angles and squared-mass differences that shape the way they change their flavor while traveling  \cite{PDG2012}. In particular, precisely known are the neutrino mixing angles of the Maki-Nakagawa-Sakata-Pontecorvo (MNSP) matrix that relates the interaction (flavor) to the propagation (mass) eigenstate basis \cite{Maki:1962mu}, the squared-mass difference values and one sign, while the other sign remains unknown (the hierarchy problem). 
Addressing the question of the value of the (Dirac or Majorana) CP violating phases 
is one of the major future goals, jointly with the
determination of the neutrino absolute mass and of the  neutrino (Dirac or Majorana) nature \cite{Giunti:2007ry}. 

Astrophysical and cosmological environments produce copious amounts of neutrinos. Therefore, a precise knowledge of neutrino flavor conversion in media is required, to assess e.g. the neutrino impact on the supernova dynamics and on (stellar or primordial) nucleosynthesis processes, to interpret the signal associated with solar neutrinos, to predict the one produced by core-collapse supernovae, or to understand how neutrinos change their flavor while traversing the Earth. It is experimentally established that the origin of the solar neutrino deficit is a resonant flavor conversion induced by neutrinos interacting with matter. This is the well known Mikheev-Smirnov-Wolfenstein (MSW) effect \cite{Wolfenstein1977,M&S1986}. More precisely, the MSW effect produces a deficit of the high energy ($^{8}$B) neutrinos, while averaged vacuum oscillations account for the one of low energy ($^{7}$Be, $pp$ and $pep$) $\nu$. 
While the MSW is the reference phenomenon to understand how neutrinos change their flavor in media, various other phenomena impacting flavor conversion can occur, depending on the specific environment under consideration. 
For example, Pantaleone has first pointed out the presence of a non-linear refractive index due to the neutrino interaction with other neutrinos whenever the neutrino number density is large \cite{Pantaleone:1992eq}.
This is relevant for neutrino evolution in the early universe, in core-collapse supernovae, and for low energy neutrinos produced in accretion-disk black-hole scenarios. Indeed,
simulations implementing the neutrino-neutrino interaction
show the emergence of new phenomena as pointed out by Samuel \cite{Samuel:1993uw}, that can be interpreted as the synchronization of effective spins \cite{Pastor:2001iu}, a flavor \cite{Duan:2007mv} or gyroscopic pendulum \cite{Hannestad:2006nj},
or a magnetic resonance phenomenon \cite{Galais:2011gh} (see e.g. \cite{Duan:2010bg} for a review).
Features associated with the explosion dynamics of a core-collapse supernova, such as the presence of shock-waves and of turbulence, produce new interference phenomena, like multiple MSW resonances \cite{Dasgupta:2005wn} and eventually depolarization \cite{Fogli:2006xy,Kneller:2010sc}. 
The spectral and location changes of the supernova neutrino fluxes impact supernova observations and are currently being investigated.
In the cosmological context, neutrino flavor conversion is also important at the epoch of Big-bang nucleosynthesis (see e.g. \cite{Dolgov:2002wy,Iocco:2008va} for a review). 
Numerous works have investigated for example the effects of mixings among active flavors  \cite{Dolgov:2002ab,Abazajian:2002qx,Mangano:2010ei}, with a possible non-zero leptonic CP violating phase \cite{Gava:2010kz, Mirizzi:2012we} or between active and sterile neutrinos  \cite{Dolgov:2003sg,Abazajian:2004aj,Hannestad:2012ky,Mirizzi:2012we}  on the primordial element abundance(s).

Numerous works have aimed at formulating theoretically the equations of motion that describe the evolution of particles with mixings in a medium.
The MSW effect is usually accounted for, by using an effective Hamiltonian
that is linear in the weak coupling constant, and depends upon the matter number density \cite{Giunti:2007ry}. In \cite{Pantaleone:1992eq} a generalization of the neutrino evolution equations is made in a similar way, to implement interaction of neutrinos with themselves. Pantaleone already emphasizes the complexity inherent with the non-linearity and the many-body character of the problem. 
Ref.\cite{Samuel:1993uw} has given a mean-field equation including such interaction terms. 
Reference \cite{Dolgov:1980cq} has first derived evolution equations for neutrino density matrices including such interaction terms.
Refs.\cite{Raffelt:1992uj} and \cite{Sigl:1992fn}
have derived neutrino evolution equations beyond the mean-field approximation and including two-body collision terms with the "molecular chaos" assumption that neglects the building up of correlations in the collision term. Such Boltzmann equations are formulated in terms of matrices of neutrino densities. 
Ref.\cite{McKellar:1992ja} has also derived the neutrino Boltzmann evolution equation using first quantization, having in mind the case of the early universe. Ref.\cite{Sawyer:2005jk} has generalized the 
equations of \cite{Sigl:1992fn} to the three flavor case (without the collision terms) and made the angular dependence more explicit. 
Many-body aspects, and the possible breakdown of the one-body description, have been discussed in Refs.\cite{Friedland:2003dv,Friedland:2003eh,Friedland:2006ke}, using the spin-spin analogy in simplified models. 
Liouville equations for neutrino distribution matrices are derived in \cite{Cardall:2007zw}. An algebraic approach to the neutrino propagation in media is given in \cite{Balantekin:2006tg} and the evolution of the many-body problem is formulated as a coherent-state path integral. This allows one, in particular, to calculate corrections to the mean-field equations as a determinant coming from the path integral \cite{Balantekin:2006tg}. The algebraic based formulation is further employed in \cite{Pehlivan:2011hp}, where the neutrino Hamiltonian of the many-body system (with mixings and the neutrino-neutrino contribution but without the matter term)  is put in connection with the (reduced) BCS pairing Hamiltonian describing superconductivity. It is pointed out that the corresponding constants of motion show the exact solvability of the problem. 

The present work formulates the problem of the neutrino evolution, in terms of reduced density matrices, using the Bogoliubov-Born-Green-Kirkwood-Yvon (BBGKY) hierarchy  \cite{Bogoliubov,Born-Green,Kirkwood,Yvon}. This replaces the Liouville Von-Neumann equation for the many-body density matrix describing the full many-body problem, by an (unclosed) set of coupled integro-differential equations for the reduced density matrices. 
Using this theoretical framework,
we derive the equations of motion for a system of neutrinos, traveling in a medium made of ordinary matter, and interacting with each other.
First we show that truncating the hierarchy at lowest order produces the 
mean-field equations for the one-body density matrix, commonly used in the literature, to implement the neutrino interaction with matter and with neutrinos. Next, we focus on the neutrino evolution description beyond the mean-field approximation and include for the first time 
neutrino-antineutrino pairing correlations to the two-body density matrix. 
Such bilinear products have been neglected so far, based upon the argument that their expectation value over free states typically oscillate fast around zero.
However, since the neutrino evolution equations are often non-linear, it is worthwhile to investigate their possible impact on neutrino flavor evolution in a medium.
We show that the inclusion of neutrino-antineutrino pairing correlations leads to extended time-dependent 
mean-field equations both for the normal and for an abnormal neutrino density matrix. 
Finally we conclude by discussing the possible implications of these contributions for neutrino flavor conversion in the astrophysical
and cosmological environments.

The manuscript is organised as follows. Section II presents the theoretical framework of the BBGKY hierarchy, its lowest order truncation that furnishes the mean-field approximation and the evolution equation for the two-body correlation function. In Section III we rederive the mean-field neutrino evolution equations including both the coupling to matter (the MSW contribution) and to neutrino themselves (the $\nu\nu$ interaction term). We discuss the relationship with the equations commonly used in the literature. Section IV introduces the contribution from $\nu\bar{\nu}$ pairing correlations to the two-body correlation function. Our extended mean-field equations for the normal and abnormal neutrino density matrices, involving the normal and pairing mean-fields, are presented. Section V includes a discussion and our conclusions.

\section{The theoretical framework}
\subsection{The Bogoliubov-Born-Green-Kirkwood-Yvon (BBGKY) hierarchy} \label{sec:bbgky} 
\noindent
In numerous contexts one is interested in determining the dynamics of a system made up of N interacting particles, such a gas of weakly interacting neutrinos, or an ensemble of strongly interacting nucleons in a nucleus or in a collision among nuclei.
The Hamiltonian for such a system of N-particles, interacting through a two-body interaction, reads
\begin{equation}\label{e:H}
\hat{H} = \sum_k \hat{H}_0(k) + \sum_{k < k'} \hat{V}(k,k')~\, 
\end{equation}
comprising the one-body $ \hat{H}_0$
kinetic and the two-body $ \hat{V}$ interaction terms. 
The $k,k'$ indices run over  single-particle quantum states, identified by single-particle properties like momentum, flavor, helicity, isospin, etc...  The system's evolution is determined by solving the Schr\"odinger equation for the many-body quantum state $| \psi (t) \rangle$ in case of a pure state or, more generally, the Liouville Von-Neumann equation for the many-body density matrix $ \hat{D}$:
\beq\label{e:LvN}
i {{d \hat{D}}\over{dt}} = [ \hat{H}, \hat{D}], 
\eeq
with $ \hat{H}$ given by Eq.(\ref{e:H}) (here we take $\hbar=c=1$).
In the BBGKY hierarchy\footnote{See e.g. Ref.\cite{Simenel:2008mh}. } theoretical framework \cite{Yvon,Kirkwood,Born-Green,Bogoliubov}, one introduces the $s$-reduced density matrix $ \hat{\rho}^{1 \ldots s}$ defined as
\beq\label{e:rhos}
\ \hat{\rho}^{1 \ldots s} = {N! \over {(N-s)!}} tr_{s+1 \ldots N}{ \hat{D}},
\eeq
$tr_{s+1}$ indicating that we are tracing over the $s$+1 particle, 
and replaces Eq.(\ref{e:LvN}) for $\hat{D}$, by an unclosed equation for $ \hat{\rho}^{1 \ldots s}$:
\beq\label{e:bbgky}
i {{d\rho_{1 \ldots s}}\over{dt}} = [H^{(s)},\rho_{1 \ldots s}]+tr_{s+1}[V^{(1\ldots s)}_{s+1},\rho_{1 \ldots s+1}]  ,
\eeq
where\footnote{Note that we do not write explicitly "hat" over the creation and annihilation operators through the whole manuscript, not to overburden the text. }  
\beq\label{e:rhocomp}
\rho_{1 \ldots s} = \langle a^{\dagger}_s \ldots a^{\dagger}_1 a_1 \ldots  a_s \rangle 
\eeq
denotes the s-body density matrix components\footnote{Note that we denote with $ \hat{\rho}^{1 \ldots s}$ the operators, while we indicate the density matrix components with $\rho_{1 \ldots s}$  or $\rho (1 \ldots s)$.}.
The $a_s$ and $a^{\dagger}_s$ correspond to the particle annihilation and creation operators for a particle in the quantum state $s$, respectively. 
In particular, the one-body and two-body matrix elements components 
are:
\beq\label{e:rho1c}
\rho_{1} = \langle a^{\dagger}_1 a_1\rangle,
\eeq
\beq\label{e:rho2c}
\rho_{12} = \langle a^{\dagger}_2 a^{\dagger}_1 a_1 a_2 \rangle,
\eeq
In Eq.(\ref{e:bbgky}) $H^{(s)}$ is the Hamiltonian of the sub-system of $s$ interacting particles, while $V^{(1\ldots s)}_{s+1}= \sum_{k} V(k,s+1)$ with $k$=1\ldots $s$. This equation is unclosed since
the $\rho_{1 \ldots s}$ evolution is coupled to the one of the ($s$+1)-reduced density $\rho_{1 \ldots s+1}$, via the two-body interaction. 
The BBGKY hierarchy (\ref{e:bbgky}) can easily be deduced by applying successive traces to Eq.(\ref{e:LvN}) and using the property
\beq\label{e:rhoprop}
\rho_{1 \ldots s} = {1 \over {N-s}} tr_{s+1} \rho_{1 \ldots s+1} = {{N!} \over{(N-s)!}} tr_{s+1, \ldots N} D,
\eeq

More explicitly Eq.(\ref{e:bbgky}) can be written as a hierarchy of equations of motion for
the $1$-body $\rho_{1}$ to the $s$-reduced $\rho_{1 \ldots s}$ density matrix components\footnote{Note that from now on we will denote $d/dt$ with a dot.}:
\begin{equation}
\left\{ 
\begin{array}{lcl} 
i \dot{\rho}_{1} & = & [H_0 (1),\rho_{1} ]+tr_{2}[V(1,2),\rho_{12}]  \\ \label{e:hierarchy}
i \dot{\rho}_{12}& = & [H_0(1) + H_0(2) + V(1,2),\rho_{12} ]    \\
&& + tr_{3}[V(1,3)+V(2,3),\rho_{123}]   \\
\vdots   \\ 
i \dot{\rho}_{1 \ldots s} & = & [\sum_{k=1}^{s} H_0(k) + \sum_{k'>k=1}^{s} V (k,k'),\rho_{1\ldots s} ]  \\
 && + \sum_{k=1}^{s} tr_{s+1}[ V(k,s+1),\rho_{1 \ldots s+1}] 
\end{array} \right .
\end{equation}
Solving Eq.(\ref{e:hierarchy}) is completely equivalent to determining the exact evolution for $\hat{D}$ Eq.(\ref{e:LvN}). The advantage of the BBGKY framework is that it furnishes a hierarchy of evolution equations for the reduced density matrices of increasing order, so that one can test different approximations, by going at a higher truncation level in the hierarchy.

\subsubsection{The mean-field approximation for the evolution equations}
\noindent
Let us consider the first equation of the BBGKY hierarchy:
\beq\label{e:mfc}
i \dot{\rho}_{1} =  [H_0 (1),\rho_{1} ]+tr_{2}[V(1,2),\rho_{12}],
\eeq
One can separate the correlated from the uncorrelated\footnote{We will also use "linked" and "unlinked" to denote the correlated and uncorrelated contributions, respectively.} contribution of the two-body density matrix\footnote{Note that here $\rho_{2}=\rho_{1}(2)$ is the one-body density matrix associated with particle 2.}:
\beq\label{e:rho2}
\rho_{12} = \rho_{1}\rho_{2} + c_{12},
\eeq
where $c_{12}$ is the two-body correlation function. 
If one deals with identical fermionic particles the uncorrelated contribution $ \rho_{1}\rho_{2}$ has to be replaced by 
$ \rho_{1}\rho_{2} (1 - P_{12})$, with $P_{12}$ being the operator that exchanges particle 1 with particle 2, to properly account for the antisymmetrization.
Inserting Eq.(\ref{e:rho2}) in Eq.(\ref{e:mfc}), one obtains
\beq\label{e:mf}
i \dot{\rho}_{1}  = [H_0 (1),\rho_{1} ] + tr_{2}[V(1,2),\rho_{1}\rho_{2}] + tr_{2}[V(1,2),c_{12}]  
\eeq
Here, since no approximation is made, the dynamical equation for the 
one-body density is exact. 

Now, neglecting the correlated contribution to the two-body correlation function, one gets
\beq\label{e:mf}
i \dot{\rho}_{1}  =  [H_0 (1),\rho_{1} ] + tr_{2}[V(1,2),\rho_{1}\rho_{2}] 
\eeq
or, equivalently:
\beq\label{e:tdhf1}
i \dot{\rho}_{1}  =  [h_1(\rho),\rho_{1} ] 
\eeq
with $h_1(\rho) = H_0 (1) + \Gamma_1(\rho)$ and $ \Gamma_1(\rho)= tr_{2}(V(1,2)\rho_{2})$ being the mean-field acting on particle 1.
This is the so-called mean-field approximation. 

Writing such an equation more explicitly, it reads
\beq\label{e:tdhf2}
i \dot{\rho}_{1,ij} - [H_0(1) + \Gamma_1(\rho), \rho_1]_{ij} = 0
\eeq
with
\beq\label{e:Gamma}
 \Gamma_{1,ij}(\rho) = \sum_{mn} v_{(im,jn)}\rho_{2,nm}.
\eeq
The mean-field potential is built up from a complete set of one-body density matrix components for particle 2 $\rho_{2,nm}$, each contributing with the matrix element\footnote{Note that in case of identical particles the matrix elements are antisymmetrised, i.e. $\tilde{v}_{(im,jn)} = \langle im | V_{12} | jn \rangle - \langle im | V_{12} | nj \rangle$.} $v_{(im,jn)} = \langle im | V_{12} | jn \rangle$, 
with $jn $ ($im$) incoming (outgoing) single-particle states.
From Eq.(\ref{e:Gamma}) one can see the dependence of our mean-field on 
the one-body density associated with particle 2, while in some cases the interaction itself might also have an explicit dependence on $\rho$.
To solve Eq.(\ref{e:tdhf2}) one has to assign 
the state of the many-body system at initial time, which can be either a correlated state, or a product of independent single-particle states. In the latter case,
the condition inherent to Eq.(\ref{e:tdhf2}), i.e.  $c_{12} =0$,  ensures that it stays as such at any time. 

It is worthwhile to mention that a first-order evolution equation for the one-body density matrix associated with a given $\hat{D}$
 can also be obtained by applying the Ehrenfest theorem:
\beq\label{e:mf2}
i \dot{\rho}_{1,ij} = \langle  [a^{\dagger}_j a_i, \hat{H} ]\rangle,
\eeq
with $\rho_{1,ij} = \langle a^{\dagger}_j a_i \rangle$ and $\hat{H}$ given by Eq.(\ref{e:H}) (in second quantization). 
In particular, Eq.(\ref{e:tdhf2}) is recovered when neglecting the correlated contribution to the two-body density matrix  \cite{Ring}.

\subsubsection{Beyond the mean-field approximation}
\noindent
Our main goal will be to discuss contributions beyond the mean-field approximation given by Eq.(\ref{e:tdhf1}), to the evolution equations for a system of relativistic neutrinos that interact among themselves and with matter. 
To this aim a useful reformulation of  Eq.(\ref{e:hierarchy}) is given by a hierarchy of evolution equations for the correlation functions (details of the demonstration can be found in  Ref.\cite{wang85}), where only linked terms are shown to remain. 
Such a reformulation has the advantage that higher-order contributions are expected to decrease with increasing rank \cite{wang85}. 
In this context, the mean-field equation (\ref{e:tdhf1}) is unchanged; while one gets for the two-body correlation function \cite{wang85}
\begin{equation}
\begin{array}{lcl} 
i \dot{c}_{12} & = & [h_1(\rho) + h_2(\rho), c_{12}]  \\
&& + (1-\rho_1)(1-\rho_2)V(1,2)\rho_{1}\rho_{2}(1-P_{12}) \\
&&- (1-P_{12})\rho_{1}\rho_{2}V(1,2)(1-\rho_1)(1-\rho_2) \\
&& +  (1-\rho_1 -\rho_2)V(1,2) c_{12} - c_{12}V(1,2)(1-\rho_1 -\rho_2)  \\
&&+ tr_3[V(1,3),(1-P_{13})\rho_1c_{23}(1-P_{12})] \\
&&+ tr_3[V(2,3),(1-P_{23})\rho_2c_{13}(1-P_{12})] \label{e:wcc12}
\end{array}
\end{equation}
with $h_1(\rho)$ ($h_2(\rho)$) the mean-fields acting on particles 1 (2) respectively and $P_{13}$ ($P_{23}$) is the operator that exchange particle 1 (2) with 3
(see Appendix A for an explicit formulation of Eq.(\ref{e:wcc12})).
Such an equation contains three main contributions coming from two-body interactions. Retaining the second and third term on the {\it r.h.s.} of Eq.(\ref{e:wcc12}), and making the "molecular chaos" assumption, that the build up of correlations due to collisions is negligible, one gets a collision term with incoming and outgoing particles described by free particle states (see e.g. \cite{lacroix}). Such a term gives rise to a Boltzmann equation\footnote{Note that  a Boltzmann equation for a system of neutrinos and antineutrinos is derived in \cite{Raffelt:1992uj,Sigl:1992fn} and in \cite{McKellar:1992ja}. Such evolution equations include a collision term 
as required in the context of the early Universe. In core-collapse supernova simulations, full transport equations for neutrinos (but without the inclusion of mixings) are usually employed in the dense region where neutrinos are trapped (see e.g. \cite{Liebendoerfer:2003es,Mezzacappa:2005ju,Woosley:2005,Kotake:2005zn,Janka:2012}).}.
The $(1-\rho_1)(1-\rho_2)$ factor\footnote{This factor as well as the $(1-\rho_1 -\rho_2)$ one come from the linked contribution of the trace term over the third particle.}  ensures the appropriate statistics (no contribution if the final single-particle states are already occupied). The fourth and fifth terms on the {\it r.h.s.} of Eq.(\ref{e:wcc12}) 
implements contributions from the correlated part of the two-body correlation function. 
It is on this term that we will focus later on. In particular we will consider cases involving both particles and antiparticles. In such systems, one might have possible contributions from the expectation values of
the product operators of the type $a_k b_l $ and $a^{\dagger}_k b^{\dagger}_l $.
Such bilinear products include particle-antiparticle correlations that can be seen as correlations of the pairing type. These terms have been neglected so far, based on the argument their expectation value (over free states) typically oscillate fast around zero, $\langle a^{\dagger}(\vec{p},t) b^{\dagger} (\vec{p}\,',t)\rangle $
(see e.g. \cite{Sigl:1992fn,Cardall:2007zw}). 
Finally, the three-body terms give the contribution from the two-body interaction among three particles, obtained by tracing over the third particle. We will neglect these higher order correlations here.
Since, in this work, we focus on the inclusion of $\nu\bar{\nu}$ contributions to $c_{12}$, we will not consider the collision and the three-body terms.
Our evolution equation for the two-body correlation function is
\begin{align}\label{e:finc}
i \dot{c}_{12} & = [h_1(\rho) + h_2(\rho), c_{12}] \\ \nonumber
 &  \quad +(1-\rho_1 -\rho_2)V(1,2) c_{12} - c_{12}V(1,2)(1-\rho_1 -\rho_2) \\ \nonumber
\end{align}

\subsection{The application to neutrinos}
\noindent
We are here mainly interested in discussing the neutrino evolution and flavor conversion when neutrinos propagate in an astrophysical environment, or in the early universe. While the BBGKY hierarchy is generally employed for systems of interacting particles without mixings, we here consider that the density matrices in Eq.(\ref{e:bbgky}) are associated with mixed particles. 
Note that a "matrix of densities", generalizing the usual occupation numbers, is commonly used in the literature (see e.g. \cite{Raffelt:1992uj,Sigl:1992fn}). 
More explicitly, in the three flavors case the neutrino density matrix  reads
\beq\label{e:rho}
\rho_{\nu} = \left(
\begin{array}{ccc} \langle a^{\dagger}_{\nu_{\alpha},i} a_{\nu_{\alpha},i} \rangle &  \langle a^{\dagger}_{\nu_{\beta},j} a_{\nu_{\alpha},i} \rangle  
&  \langle a^{\dagger}_{\nu_{\gamma},k} a_{\nu_{\alpha},i} \rangle \\
\langle a^{\dagger}_{\nu_{\alpha},i} a_{\nu_{\beta},j} \rangle
& \langle a^{\dagger}_{\nu_{\beta},j} a_{\nu_{\beta},j} \rangle  & \langle a^{\dagger}_{\nu_{\gamma},k} a_{\nu_{\beta},j} \rangle \\ 
 \langle a^{\dagger}_{\nu_{\alpha},i} a_{\nu_{\gamma},k} \rangle & \langle a^{\dagger}_{\nu_{\beta},j} a_{\nu_{\gamma},k} \rangle  & \langle a^{\dagger}_{\nu_{\gamma},k} a_{\nu_{\gamma},k} \rangle
 \end{array}
\right).
\eeq
The off-diagonal (or coherent) terms are non-zero to encode the presence of neutrino mixings.
The neutrino occupation number for a given $\nu_{{\alpha},i}$ flavor 
state is given by the diagonal element of the density matrix $\langle a^{\dagger}_{\nu_{\alpha},i} a_{\nu_{\alpha},i} \rangle$, with $N_{\nu_{\alpha}} = \sum_i \langle a^{\dagger}_{\nu_{\alpha},i} a_{\nu_{\alpha},i} \rangle $ the total occupation number.
The definition in Eq.(\ref{e:rho}) can easily  be extended to the case of an arbitrary neutrino families,
to account for the presence of both sterile and active neutrinos. 

Since the systems we are interested in involve both particles and antiparticles, 
one introduces a density matrix $\bar{\rho}_{\nu}$, in a way analogous to Eq.(\ref{e:rho}), but replacing the particle operators $a^{\dagger},a$ with the
antiparticle $b^{\dagger},b$ ones \cite{Sigl:1992fn}. For the sake of clarity concerning the convention used in the present work, we give its explicit expression:
\beq\label{e:rhoanu}
\bar{\rho}_{\nu} = \left(
\begin{array}{ccc} \langle b^{\dagger}_{\nu_{\alpha},i} b_{\nu_{\alpha},i} \rangle &  \langle b^{\dagger}_{\nu_{\beta},j} b_{\nu_{\alpha},i} \rangle  
&  \langle b^{\dagger}_{\nu_{\gamma},k} b_{\nu_{\alpha},i} \rangle \\
\langle b^{\dagger}_{\nu_{\alpha},i} b_{\nu_{\beta},j} \rangle
& \langle b^{\dagger}_{\nu_{\beta,j}} b_{\nu_{\beta},j} \rangle  & \langle b^{\dagger}_{\nu_{\gamma},k} b_{\nu_{\beta},j} \rangle \\ 
 \langle b^{\dagger}_{\nu_{\alpha},i} b_{\nu_{\gamma},k} \rangle & \langle b^{\dagger}_{\nu_{\beta},j} b_{\nu_{\gamma},k} \rangle  & \langle b^{\dagger}_{\nu_{\gamma},k} b_{\nu_{\gamma},k} \rangle
 \end{array}
\right).
\eeq

To implement the antiparticle degrees of freedom, we consider the usual
expansion of the neutrino fields in the Schr\"odinger picture
\begin{align}\label{e:fields}
\phi (\vec{x})
&  = \sum_h \int {d^3 \vec{p} \over{(2 \pi)^3} 2 E_{p}} [a(\vec{p},h)u_{\vec{p},h} e^{i \vec{p} \cdot \vec{x}} \\ \nonumber
& \quad \quad \quad \quad \quad + b^{\dagger}(\vec{p},h) v_{\vec{p},h} e^{-i \vec{p} \cdot \vec{x}}],
\end{align}
The summation is over the helicity $h$ states, the integration over the momenta $\vec{p}$ of the (anti)particles and  $u_{\vec{p},h}$ ($v_{\vec{p},h}$) are the usual Dirac spinors, $E_{p}$ being neutrino energy. Note that in such an expression the
particle creation and annihilation operators are associated with states of
a given mass for which 
the usual anticommutation rules are properly defined. In the present manuscript we employ\footnote{Similar relations hold for the antiparticle annihilation $ b(\vec{p},h)$ and creation 
$b^{\dagger}(\vec{p}\,',h')$ operators.} 
\beq\label{e:commutators1}
\{ a(\vec{p},h), a^{\dagger}(\vec{p}\,',h') \} 
 = (2 \pi)^3 2 E_{p}\delta^3(\vec{p} - \vec{p}\,')\delta_{hh'}
\eeq 
and
\beq\label{e:commutators2}
\{ a(\vec{p},h), a(\vec{p}\,',h') \} = 0.
\eeq 
The single-particle states associated with neutrino mass eigenstates are
\beq\label{e:sp}
| m \rangle = a^{\dagger}_m | \rangle
\eeq
with $| \rangle$ being the vacuum state defined by $a_m | \rangle = 0$. 
The flavor eigenstates are related to the mass eigenstates through
$| \nu_{\alpha} \rangle= \sum_{i} U^*_{\alpha i} |\nu_i \rangle$ ($i$ and $\alpha$ correspond to an arbitrary number of neutrino families), where $U$ is the Maki-Nakagawa-Sakata-Pontecorvo (MNSP) unitary matrix. In three flavors, the MNSP matrix depends upon three neutrino mixing angles that have been measured, one Dirac and two Majorana unknown phases \cite{PDG2012}. Since the appropriate anticommutation rules only hold for the mass eigenstates operators, 
Eqs.(\ref{e:commutators1}--\ref{e:commutators2}) require special attention (see e.g. Ref.\cite{Cardall:2007zw} for a discussion). This subtlety is sometimes avoided by writing the fields Eq.(\ref{e:fields}) for massless neutrinos, as done e.g.\cite{Sigl:1992fn}. As discussed for example in Refs.\cite{Cardall:2007zw,Giunti:1991cb}, it is not possible to rigorously build up a Fock space for flavor states since the neutrino flavor creation and annihilation operators do not satisfy the canonical anticommutation rules 
Eqs.(\ref{e:commutators1}-\ref{e:commutators2}). Indeed, as shown in Ref.\cite{Giunti:1991cb},
one can define an approximate Fock space by introducing neutrino "weak-particle states" that depend upon the specific weak process under consideration. However, in the limit of relativistic neutrinos, the anticommutation relations (\ref{e:commutators1}-\ref{e:commutators2}) also approximately hold for flavor states.
Since this is a good approximation for our cases of interest (solar and supernova neutrinos, cosmological neutrinos at the epoch of big-bang nucleosynthesis), we make the assumption that an approximate Fock space for our flavor states can be built.
It is worthwhile to mention that extending our results without making this approximation does not introduce extra conceptual difficulties. 
In much the same way as done in Ref.\cite{Giunti:1991cb} for the muon decay case,
but in the  mean-fields expressions,
one should retain an explicit dependence on the mixing matrix elements in the  
calculation of the interaction matrix elements,
depending on the specific weak process under consideration\footnote{To implement this correction, one should keep in mind that we deal with the process amplitudes, and not amplitude squares as in \cite{Giunti:1991cb}.} .

Finally, it is worthwhile to remind that, while the BBGKY theoretical framework  is based upon the density matrix formalism, the evolution equations (\ref{e:hierarchy}) 
can be formulated as a hierarchy for many-body Greens' functions as
done in  Ref.\cite{Wang:1994xc}. In particular, in the equal time limit,
when only linked contributions are retained, these two formalisms are completely equivalent.
Note that while historically BBGKY was developed to describe the evolution of nonrelativistic systems of N-particles, the hierarchy applies to a system of relativistic particles (as of interest here). In this case Eqs.(9) are replaced by an infinite  set of equations.

\section{Neutrinos evolving in a medium in the mean-field approximation}
\noindent
We now take the example of neutrinos interacting with 
the electrons, protons and neutrons composing a medium
to show how the formalism presented in Section II can be 
used to rigorously derive well known neutrino evolution equations. 
We just sketch the main lines of the derivation\footnote{A different  
derivation of the mean-field equations accounting for the neutrino interaction with a medium is given e.g. in \cite{Giunti:2007ry}.}.
 The assumption that is usually made is that our system of
         neutrinos interacting with a medium can be described at lowest order as a  system 
         of independent particles, so that implicitly the 
         problem reduces 
         to following the evolution of a single-particle at a time and calculating 
the evolution of the associated one-body density matrix.  

Let us consider the BBGKY hierarchy truncated in the mean-field approximation given by Eq.(\ref{e:tdhf1}). 
The Hamiltonian Eq.(\ref{e:H}) for our case of interest  in the flavor basis reads

\beq\label{e:Hf}
H^f = UH_mU^{\dagger} + H_{int},
\eeq
where $U$ is the MNSP unitary matrix relating the neutrino flavor basis to the mass eigenstate  basis (with eigenenergies $E_i$). The first $H_m = diag(E_i)$ contribution is the propagation term, while the second one $H_{int}$ corresponds to the two-body interaction between a neutrino and another particle. 

\subsection{The mean-field associated with neutrino interaction with matter (MSW case)}
\noindent

\begin{figure}[!]
\includegraphics[scale=0.35]{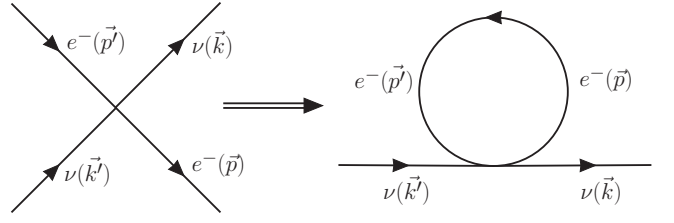}
\caption{Neutrino interaction with electrons and the corresponding mean-field associated with the electron background. 
In the evolution equations,  the mean-field acting on a single neutrino state is build up from the summation of the electron single-particle states (see Eqs.(\ref{e:Gamma}), (\ref{e:Gmsw}) and (\ref{e:msw})).}
\label{fig:mfe}
\end{figure}

To follow the one-body density matrix evolution given by Eq.(\ref{e:mf2}), one needs to determine the mean-field (\ref{e:Gamma}) created by the background particles and acting on the "test" neutrino. 
	 The interaction term $H_{int}$ corresponds to the charged- or neutral-current Hamiltonian describing the neutrino interaction with the medium. We take the example of the charged-current interaction on electrons, where $H_{int}$ is given by\footnote{The effective low energy approximation is sufficient for the applications envisaged.} 
\beq\label{e:Hcc}
H_{CC} = {G_F \over{\sqrt{2}}} \int d^3 \vec{x} [\bar{\phi}_e\gamma_{\mu}(1-\gamma_5)\phi_{\nu_e}],
[\bar{\phi}_{\nu_e}\gamma^{\mu}(1-\gamma_5)\phi_e]
\eeq
where $G_F$ is the Fermi coupling constant and 
with the fields $\phi$ given by Eq.(\ref{e:fields}).
	 This requires 
calculating the matrix elements for neutrino-electron scattering
$v^{\nu_e,e}_{im,jn} = \langle \nu_{ei} e_{m} | H_{CC} | \nu_{ej} e_n \rangle$
with the interaction Hamiltonian (\ref{e:Hcc}). 
One gets 
\begin{eqnarray}\label{e:Gmsw}
\Gamma_{\nu_e}(\rho_e)  & = & {{G_F} \over{\sqrt{2}}} \sum_{h_e,h'_e} \int  {{d^3 \vec{p}} \over{(2 \pi)^3 2 E_{p}}}
\int {{d^3 \vec{p}\,'}\over{(2 \pi)^3 2 E_{p'}}} \nonumber \\
&&(2 \pi)^3 \delta^3(\vec{p} + \vec{k} - \vec{p}\,' -\vec{k}')   \nonumber \\
&& [\bar{u}_{\nu_e} (\vec{k},h_{\nu_e})\gamma_{\mu}(1-\gamma_5) u_{\nu_e} (\vec{k}',h'_{\nu_e})]  
\nonumber \\
& & [\bar{u}_e (\vec{p},h_{e})\gamma^{\mu}(1-\gamma_5) u_{e} (\vec{p}\,',h'_e)]  \nonumber \\
& & \langle a^{\dagger}_e (\vec{p},h) a_e (\vec{p}\,',h')\rangle.
\end{eqnarray}
The summation over the $m,n$ single-particle states in Eq.(\ref{e:Gamma}) becomes here a sum over 
the electron helicity states and an integration over momenta (Figure \ref{fig:mfe}).

The key quantity to define is the expectation value of the 
$a^{\dagger}_e (\vec{p},h) a_e (\vec{p}\,',h')$ operator, over the electron background\footnote{Here we make the assumption that  the contributions from the neutrino and the electron backgrounds can be factorized in $\hat{D}$, i.e. that modifications of the electron background coming from the interaction with neutrinos can be neglected. } 
 characteristic of the considered medium\footnote{Note that in this and in the following section, we denote the single-particle density $\rho_{1,kl}$ as $\rho_{kl}$ to simplify notations. If, a particle is
identified by helicity and momentum, then $\rho_{kl}$ reads 
$\rho^e_{\vec{p}h,\vec{p}\,'h'}$ for an electron background, for example.}:
\beq\label{e:rhoe}
\rho^e_{\vec{p}\,'h',\vec{p}h} \equiv  \langle  a^{\dagger}_e (\vec{p},h) 
a_e (\vec{p}\,',h')  \rangle.
\eeq
The assumption of a homogeneous and unpolarized medium corresponds to:
\beq\label{e:rhoeh}
\rho^e_{\vec{p}\,'h',\vec{p}h} = (2 \pi)^3 2 E_{p} \delta_{hh'} \delta^3 (\vec{p} -\vec{p}\,') \rho^e_{\vec{p}}.
\eeq
 Equation (\ref{e:rhoe})
constitutes for example a good approximation for the case of the Sun.
Using Eqs.(\ref{e:rhoe}-\ref{e:rhoeh}) the total electron number is
\begin{eqnarray}\label{e:enumber}
N_e  & =&  2 V  \int {{d^3 \vec{p}} \over{(2 \pi)^3}} \rho^e_{\vec{p}},
\end{eqnarray}
where the factor of 2 comes from the summation over the electron helicity states, and $V$ is the quantization volume.
In Eq.(\ref{e:Gmsw}), by tracing over the spinors, implementing that neutrinos have only one possible helicity state, 
and assuming the electron background is homogeneous, 
unpolarized and isotropic, one gets for 
the mean-field Eq.(\ref{e:Gamma})
\beq\label{e:msw}
\Gamma_{\nu_e}(\rho_e) = (2 \pi)^3 2 E_{k} \delta^3(\vec{k}-\vec{k}') \sqrt{2} G_F n_e,
\eeq
where $n_e = N_e/V $ is the electron number density.
In Eq.(\ref{e:msw}) the $\delta$-function ensures that the momentum
 of the neutrino propagating in the electron medium is unchanged, as a consequence of homogeneity. The prefactor
$(2 \pi)^3 2 E_{k}$ are present because of the chosen normalization of the
(anti)particle anticommutation relations Eqs.(\ref{e:commutators1}-\ref{e:commutators2}).
Our result (\ref{e:msw}) is the mean-field contribution to the neutrino Hamiltonian, corresponding to neutrino interaction with matter,
that is linear in the weak coupling constant and depends upon the number densities of the particles composing the medium. We find the well known low energy effective Hamiltonian (see e.g. \cite{Giunti:2007ry}) that can give rise, if neutrinos propagate adiabatically in a medium, to a resonant\footnote{The occurence of 
such a resonant phenomenon depends upon neutrino properties (energies, mixing angles,
squared-mass differences value and sign) and the specific matter number density profile for the system under 
consideration.} flavor conversion -- the Mikheev-Smirnov-Wolfenstein effect 
\cite{Wolfenstein1977,M&S1986}.

The procedure just outlined can be applied to the neutral-current $\nu_e$ scattering on electrons, to the charged- and neutral-current $\bar{\nu}_e$ scattering on electrons,
of  $\nu_e$ or $\bar{\nu}_e$ on positrons, as well as neutral-current scattering
on protons and neutrons, giving the expected results.
In particular, it is immediate to show that, when dealing with antiparticles, 
the associated mean-field $\Gamma$ 
depends upon matrix elements that involve the $b^{\dagger}$ and $b$ operators (instead of the particle operators), 
introducing a minus sign. For example, for the case of  $\bar{\nu}_e$ evolving in an electron 
background, one obtains the expected result
\beq\label{e:msw2}
\Gamma_{\bar{\nu}_e}(\rho_e) = -
(2 \pi)^3 2 E_{k} \delta^3(\vec{k}-\vec{k}') \sqrt{2} G_F n_e .
\eeq

\subsection{The mean-field associated with neutrino self-interactions}
The second case we are going to consider, in the mean-field approximation, is when 
the (anti)neutrino is evolving in a background of $\nu$ and $\bar{\nu}$. This case is of interest since recent studies have shown the important role of the neutrino-neutrino interaction for instance in a core-collapse supernova, and the variety of new flavor conversion phenomena that can arise, when implementing this contribution in the neutrino Hamiltonian (see e.g. \cite{Duan:2010bg} for a review). A series of works have discussed the neutrino evolution equation in presence of such terms
 \cite{Pantaleone:1992eq,Samuel:1993uw,Sigl:1992fn,McKellar:1992ja,Qian:1994wh,Sawyer:2005jk,Balantekin:2006tg, Pehlivan:2011hp}. Note that the role of such terms was first pointed out in the context of the early universe \cite{Kostelecky:1993yt}. Here we just sketch the derivation of these equations, following the same procedure as for the MSW case (more details are given in Appendix B). 

\begin{figure}[!]
\begin{center}
\includegraphics[scale=0.35]{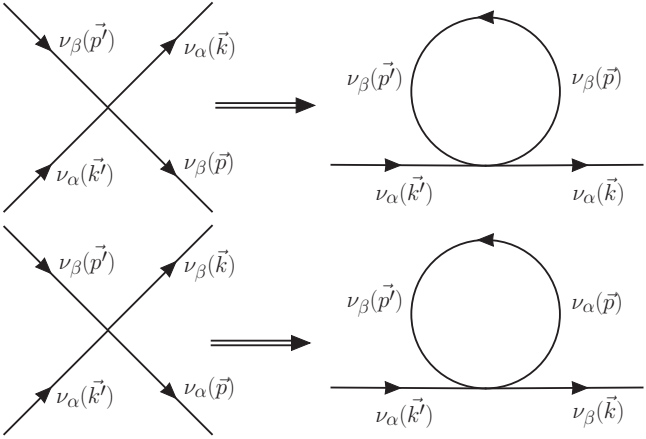}
\end{center}
\caption{The figure shows, in a pictorial way, the interaction terms and the corresponding mean-fields Eq.(\ref{e:nnu}) associated with the neutrino-neutrino interaction. The mean-field acting on a single neutrino state is build up from the summation of the single-particle states that make up the background. The two contributions correspond to the diagonal part of the mean-field $\Gamma_{\nu_{\alpha},\nu_{\alpha}}$, arising from the usual scattering terms (upper figures), and the off-diagonal part $\Gamma_{\nu_{\alpha},\nu_{\beta}}$ Eq.(\ref{e:rhonu}) (lower figures) associated with Pantaleone off-diagonal refractive index. The diagonal contribution to the mean-field  $\Gamma_{\nu_{\alpha},\nu_{\alpha}}$ has an extra term coming from neutrinos of the same flavor $\alpha$ running in the loop. Both the diagonal and the off-diagonal mean-field receives similar contributions from antineutrinos, instead of neutrinos, in the loop Eq.(\ref{e:mfnnu}). }
\label{fig:mfvv}
\end{figure}

The neutrino-neutrino mean-field depends this time on the effective low energy neutral-current Hamiltonian:
\beq\label{e:Hnc}
H_{NC} = {G_F \over{2 \sqrt{2}}} \int d^3 \vec{x} [\bar{\phi}_{\nu_e}\gamma_{\mu}(1-\gamma_5)\phi_{\nu_e}],
[\bar{\phi}_{\nu_y}\gamma^{\mu}(1-\gamma_5)\phi_{\nu_y}]
\eeq
with $\nu_y = \nu_e, \nu_{\mu}$ or $\nu_{\tau}$.
From Eq.(\ref{e:Gamma}) one gets for the mean-field $\Gamma_{\nu_{\alpha}, \nu_{\beta}} (\rho_{\nu})$ with $\nu_{\alpha},\nu_{\beta}=\nu_e,~\nu_{\mu},~\nu_{\tau}$ (or the corresponding antineutrinos):
\begin{eqnarray}\label{e:nnu}
\Gamma_{\nu_{\alpha},\nu_{\beta}}(\rho_{\nu}) 
 & = & {{G_F} \over{2\sqrt{2}}}  \int  {{d^3 \vec{p}} \over{(2 \pi)^3  2 E_{p}}}  
\int {{d^3 \vec{p}\,'}\over{(2 \pi)^3  2 E_{p'}}} \nonumber \\
&& (2 \pi)^3 \delta^3(\vec{p} + \vec{k} - \vec{p}\,' -\vec{k}')  
\nonumber \\ 
& & [\bar{u}_{\nu_{\beta}} (\vec{k},h_{\beta})\gamma_{\mu}(1-\gamma_5) u_{\nu_{\alpha}} (\vec{k}',h'_{\alpha})]  \nonumber \\
& & [\bar{u}_{\nu_{\alpha}}  (\vec{p},h_{\alpha})\gamma^{\mu}(1-\gamma_5) u_{\nu_{\beta}} (\vec{p}\,',h'_{\beta})]  \nonumber \\ 
& & \langle a^{\dagger}_{\nu_{\alpha}} (\vec{p},h_{\alpha}) a_{\nu_{\beta} } (\vec{p}\,',h'_{\beta}) \rangle
\end{eqnarray}
and requiring the expectation value 
\beq\label{e:rhonu}
\rho^{\nu_{\beta}, \nu_{\alpha}}_{\vec{p}\,'h',\vec{p}h} \equiv  
\langle  a^{\dagger}_{\nu_{\alpha}} (\vec{p},h_{\alpha}) a_{\nu_{\beta}} (\vec{p}\,',h'_{\beta}) \rangle
\eeq
correspond to a homogeneous and unpolarized system:
\beq\label{e:rhonuh}
\rho^{\nu_{\beta}, \nu_{\alpha}}_{\vec{p}\,'h',\vec{p}h}= (2 \pi)^3 2 E_{p} \delta_{hh'} \delta^3 (\vec{p} -\vec{p}\,') \rho^{\nu_{\beta}, \nu_{\alpha}}_{\vec{p}}.
\eeq
The key difference with the case of the electron background is that the quantity $\rho^{\nu_{\beta}, \nu_{\alpha}}_{\vec{p}\,'h',\vec{p}h}$
has diagonal and off-diagonal terms (Figure \ref{fig:mfvv}).
As a consequence the mean-field acting on a $\nu_e$ for example
has both diagonal  $\Gamma_{\nu_{\alpha},\nu_{\alpha}}(\rho_{\nu})$ and off-diagonal $\Gamma_{\nu_{\alpha},\nu_{\beta}}(\rho_{\nu})$ contributions.
One recognizes in the mean-field contributions that the diagonal one is associated with forward scattering without flavor exchange, while the off-diagonal contribution is associated with
forward scattering "with flavor exchange" (the Pantaleone diagram \cite{Pantaleone:1992eq}). 
We emphasize that in our derivation we do not directly start with an effective Hamiltonian
having these terms, as done e.g. in Ref.\cite{Balantekin:2006tg}. Here these contributions naturally arise when considering the weak interaction term, 
using the density matrix Eq.(\ref{e:rho}) and calculating $\Gamma $ (or, in other words, closing the loop).
By implementing that neutrinos have one helicity state only, tracing over the spinors, 
 in case the neutrino background is homogeneous and anisotropic,
one gets for the off-diagonal contribution the following expression: 
\begin{eqnarray}\label{e:nnu} 
\Gamma_{\nu_{\alpha},\nu_{\beta}}(\rho_{\nu})& = &
(2 \pi)^3 2 E_{k} 
 \delta^3(\vec{k}-\vec{k}') \nonumber \\
 && \sqrt{2} G_F \int {{d^3 \vec{p}} \over{(2 \pi)^3 }}  \rho^{\nu_{\beta},\nu_{\alpha}}_{\vec{p}} \left( 1-\hat{\vec{p}} \cdot \hat{\vec{k}}  \right)
\end{eqnarray}
with $\hat{\vec{p}} = \vec{p}/|\vec{p}|$ and $\hat{\vec{k}}   = \vec{k}/|\vec{k}| $.  A similar expression holds for the diagonal one if $\nu_{\beta} = \nu_{\alpha}$, except for an extra factor of 2. In the case of an isotropic medium the angular term $cos\theta_{\vec{k}\vec{p}} = \hat{\vec{k}} \cdot \hat{\vec{p}}$  averages out, and one recovers the same result as Eq.(\ref{e:msw}).
If the background includes antineutrinos as well, one needs to add the contribution from $\bar{\rho}^*$, but with a minus sign. By adding up the two contributions
the total neutrino-neutrino mean-field reads\footnote{Note that the contribution coming for the antineutrino density matrix is
 $\bar{\rho}^*$ if one employs the definition $\bar{\rho}_{ij}= \langle b^{\dagger}_j b_i \rangle$ (see Appendix B). If one defines such a quantity as
$\bar{\rho}_{ij}= \langle b^{\dagger}_i b_j \rangle$ then the contribution appearing in the evolution equations depends upon $\bar{\rho}$, instead of $\bar{\rho}^*$.}:
\begin{eqnarray}\label{e:mfnnu}
{\bf \Gamma}(\rho_{\nu},\bar{\rho}_{\nu})  &=& (2 \pi)^3 2 E_{k} 
 \delta^3(\vec{k}-\vec{k}') \sqrt{2} G_F \\
& &\sum_{\underline{\nu_{\alpha}}} \int {{d^3 \vec{p}} \over{(2 \pi)^3 }} (\rho_{\underline{\nu_{\alpha}},\vec{p}} - 
\bar{\rho}^*_{\underline{\nu_{\alpha}},\vec{p}} ) \left( 1- \hat{\vec{p}} \cdot \hat{\vec{k}}  \right), \nonumber 
\end{eqnarray}
where  here $\rho_{\underline{\nu_{\alpha}},\vec{p}}$ ($\bar{\rho}^*_{\underline{\nu_{\alpha}},\vec{p}} )$ stand for the density 
matrix Eq.(\ref{e:rho}) (Eq. (\ref{e:rhoanu})), $\underline{\nu_{\alpha}}$ refers to a neutrino that is initially born in the $\alpha$ flavor. In fact, one has to sum over all neutrino flavors present in the system. Note that in Eq.(\ref{e:mfnnu}) the trace term, $tr(\rho_{\underline{\nu_{\alpha}},\vec{p}}-\bar{\rho}^*_{\underline{\nu_{\alpha}},\vec{p}}$),
has been subtracted.

\subsection{Neutrino evolution equations in the mean-field approximation}
\noindent
With the results of Eqs.(\ref{e:msw}) and (\ref{e:mfnnu}), 
the mean-field equations (\ref{e:tdhf2}-\ref{e:Gamma}) for the density matrix
(\ref{e:rho})
describing the neutrino evolution in a medium becomes explicitly:
\beq\label{e:neqmf}
i \dot{\rho}  =  [h(\rho),\rho ] 
\eeq
with 
\begin{eqnarray}\label{e:neqmf2}
 h(\rho) &= &UH_mU^{\dagger} +  H_{mat}  \\
 && + \sqrt{2} G_F \sum_{\underline{\nu_{\alpha}}} \int {{d^3 \vec{p}} \over{(2 \pi)^3 }} (\rho_{\underline{\nu_{\alpha}},p} - \bar{\rho}^*_{\underline{\nu_{\alpha}},p}) \left( 1- \hat{\vec{p}} \cdot \hat{\vec{k}} \nonumber
\right) 
\end{eqnarray}
where $H_{mat} = diag(\sqrt{2} G_F n_e, 0, 0)$.
A similar equation holds if an antineutrino is traveling instead of a neutrino:
\beq\label{e:neqmfanu}
i \dot{\bar{\rho}}  =  [\bar{h}(\bar{\rho}),\bar{\rho} ] 
\eeq
with\footnote{Note that, according to our definition for $\bar{\rho}$ Eq.(\ref{e:rhoanu}), antineutrinos do not transform the same way as neutrinos under $U$.} 
\begin{eqnarray}\label{e:neqmf2anu}
 \bar{h}(\bar{\rho}) &= &U^*H_mU^{T} - H_{mat}   \\
 && - \sqrt{2} G_F \sum_{\underline{\nu_{\alpha}}} \int {{d^3 \vec{p}} \over{(2 \pi)^3 }} (\rho^*_{\underline{\nu_{\alpha}},p} - \bar{\rho}_{\underline{\nu_{\alpha}},p}) \left( 1- \hat{\vec{p}} \cdot \hat{\vec{k}} \nonumber
\right) 
\end{eqnarray}

Note that in Eq.(\ref{e:neqmf}-\ref{e:neqmf2anu}) $\rho$ and $\bar{\rho}$ have two indices in flavor, but only one in momentum as a consequence of homogeneity. In case the quantum state
 at initial time is an independent particle state,
the one-body density matrices (\ref{e:rho}-\ref{e:rhoanu}) can be replaced by single-particle one-body densities. In the neutrino case, the diagonal elements $\rho_i$ directly give the
neutrino survival probabilities $|\nu_i|^2$  $\nu_i$ being the neutrino amplitude for flavor $i$, while the off-diagonal ones $\rho_{ij}$
are the mixing terms $\nu_i\nu_{j}$ (and similarly for antineutrinos). 
We conclude here by emphasizing that the evolution equations (\ref{e:neqmf}-\ref{e:neqmf2anu}) we find are in agreement with those of 
Refs.\cite{Sigl:1992fn,Qian:1994wh,Balantekin:2006tg}, commonly used in the investigation of solar, of supernova neutrinos and of the low energy neutrinos in accretion-disks black-hole scenarios.

\section{Neutrino-antineutrino pairing correlations}
\subsection{The extended neutrino dynamical equations}
\noindent
Our main goal is to obtain the evolution equations, beyond the mean-field approximation, for a system of neutrinos and antineutrinos evolving in an environment, taking into account possible $\nu\bar{\nu}$ pairing correlations. The latter correspond to the following linked contribution to the two-body density matrix Eq.(\ref{e:rho2}):
\beq\label{e:cpair}
c_{\alpha\beta,\alpha'\beta'} 
 \approx  \langle a^{\dagger}_{\alpha'}b^{\dagger}_{\beta'}\rangle \langle  b_{\beta} a_{\alpha} \rangle \\
= \kappa^*_{\alpha'\beta'} \kappa_{\alpha\beta},
\eeq
\noindent 
where the quantities $\kappa_{\alpha\beta}$ and $\kappa^*_{\alpha'\beta'}$ are called the abnormal densities.
Imposing that the pair products conserve individual lepton numbers, 
only product of operators associated with the same flavor, such as $a_{\nu_{\alpha}}b_{\nu_{\alpha}}$,  and the corresponding hermitian conjugates, are admitted. 
Since neutrinos have mixings, one can have contributions from products involving neutrino-antineutrino pairs with different flavors. More generally, e.g.
if total lepton-number is not conserved one could have correlations associated with the pair product operators like $a_{\nu_{\alpha}} a_{\nu_{\alpha}}$  (and similarly for antineutrinos).
We note that the expectation values\footnote{Note that the spinor products are not usually modified in our cases of interest.} 
$\langle a^{\dagger}a \rangle$, $\langle b^{\dagger}b \rangle$, $\langle ba \rangle$ and $\langle a^{\dagger}b^{\dagger} \rangle$
naturally appear as components of the field correlation function $\langle \phi \bar{\phi} \rangle$.
So far, neutrino-antineutrino correlations have been neglected\footnote{Note that, in the context of
baryogenesis via leptogenesis, the authors of Ref.\cite{Fidler:2011yq}
have emphasized the role of neutrino-antineutrino correlations, in a quantum field theory approach, including the Boltzmann collision term but within a simplified neutrino model. An extension of the mean-field Eqs.(\ref{e:neqmf}-\ref{e:neqmf2anu}) including neutrino-antineutrino mixings is considered in Ref.\cite{deGouvea:2012hg} in presence of nonzero transition magnetic moments and in Ref.\cite{Sawyer:2010jk} due to the neutrino interaction with scalars.}   (see e.g. Refs.\cite{Raffelt:1992uj,Sigl:1992fn,Cardall:2007zw}).  For example, in the formulation of Ref.\cite{Cardall:2007zw},
they correspond to the
rapidly oscillating cross terms between the positive- and negative-frequency parts of the quantum density function $i\Gamma^{lm}_{ij} = \langle N \nu_{i}^{l}(y) \bar{\nu}_{j}^{m}(z) \rangle$, where the $\nu(y)$ and $\bar{\nu}(z)$ are the neutrino and antineutrino quantum field operators.

Two perspectives are possible to investigate the impact of the neutrino-antineutrino correlations on the neutrino evolution. The first is to assume that such correlations are non-zero at initial time, e.g. at the neutrino sphere in a supernova,  from previous interactions among neutrinos in the dense supernova region, where they are trapped. The second possibility is to study whether such terms can be dynamically produced through the interactions. However this 
is more demanding, since one has to retain the collision term in the
two-body correlation function Eq.(\ref{e:wcc12}) as well. In this manuscript we adopt the first perspective.
We will see below that, according to our extended evolution equations if their contribution is zero at initial time it is zero at all times. Therefore, in this case, the mean-field approximation given by Eqs.(\ref{e:neqmf}-\ref{e:neqmf2anu})  is correct (if for the considered system, the collision term can also be neglected in Eq.(\ref{e:wcc12})). This is the approximation that is usually implicitly made.

Let us now discuss how the neutrino evolution equation Eqs.(\ref{e:neqmf}-\ref{e:neqmf2anu}) have to be extended to
implement pairing correlations between $\nu$ and $\bar{\nu}$.
In this case, the evolution of the system is determined by using the first two equations of the BBGKY hierarchy Eq.(\ref{e:hierarchy}). By substituting Eq.(\ref{e:cpair}) in 
Eq.(\ref{e:finc}) one obtains the evolution equation for the abnormal density (and its complex conjugate), while from the first equation of the BBGKY hierarchy and Eq.(\ref{e:rho2}), we obtain the evolution equation for the normal density. We finally get
the extended evolution equations:
\begin{equation}
\left\{ 
\begin{array}{lcl} 
i \dot{\rho}_{ij} (1) & = & [h(1),\rho(1)]_{ij} + \sum_m (\Delta_{im} \kappa^*_{jm} - \kappa_{im} \Delta^*_{jm}  ) \\ 
i \dot{\bar{\rho}}_{kl} (2) & = & [\bar{h}(2),\bar{\rho}(2)]_{kl} + \sum_m (\Delta_{mk} \kappa^*_{ml} - \kappa_{mk} \Delta^*_{ml}  ) \\ 
i \dot{\kappa}_{ik} & = & \sum_m (h_{im}(1) \kappa_{mk} + h_{km}(2) \kappa_{im}) + \Delta_{ik}   \\ 
 &  & - \sum_m (\rho_{im} (1) \Delta_{mk} + \bar{\rho}_{km} (2) \Delta_{im} ) \label{e:tddm}
\end{array} \right. 
\end{equation}
where here the indices $i,j$ stand for $\nu_{\alpha}, \nu_{\beta}$; $k, l$ for  $\bar{\nu}_{\alpha}, \bar{\nu}_{\beta}$  with $\alpha, \beta $ that vary over  the different electron, muon and tau flavor states.
For the sake of clarity, in Eqs.(\ref{e:tddm}) we show explicitly the dependence on particle 1 and particle 2 of the quantities. Obviously, in our extended equations one has to determine the evolution of the normal density associated with a neutrino (particle 1) and an antineutrino (particle 2) to determine the neutrino-antineutrino pair evolution in $\kappa$ and $\kappa^*$.
We emphasize that, the new evolution equations Eq.(\ref{e:tddm})  can also be derived by using the Ehrenfest theorem and determining the first-order evolution equations for the abnormal density:
\beq\label{e:k2}
i \dot{\kappa}_{ik} = \langle [b_k a_i, \hat{H} ] \rangle
\eeq
and for the normal one using Eq.(\ref{e:mf2}).

The expression for the abnormal mean-field is
\beq\label{e:abpair}
\Delta_{ik} = \sum_{jl} v_{(ik,jl)} \kappa_{jl}
\eeq
and of its complex conjugate
\beq\label{e:abpairc}
\Delta^*_{ik} = \sum_{jl} v_{(jl,ik)} \kappa^*_{jl}.
\eeq
One can see that such expressions are analogous to the one for the mean-field $\Gamma$ Eq.(\ref{e:Gamma}), but with
the abnormal density $\kappa$ replacing the normal density $\rho$, and by summing over the initial (or final) single-particle states instead of over a final and initial single-particle state.
\begin{figure}[!]
\begin{center}
\includegraphics[scale=0.38]{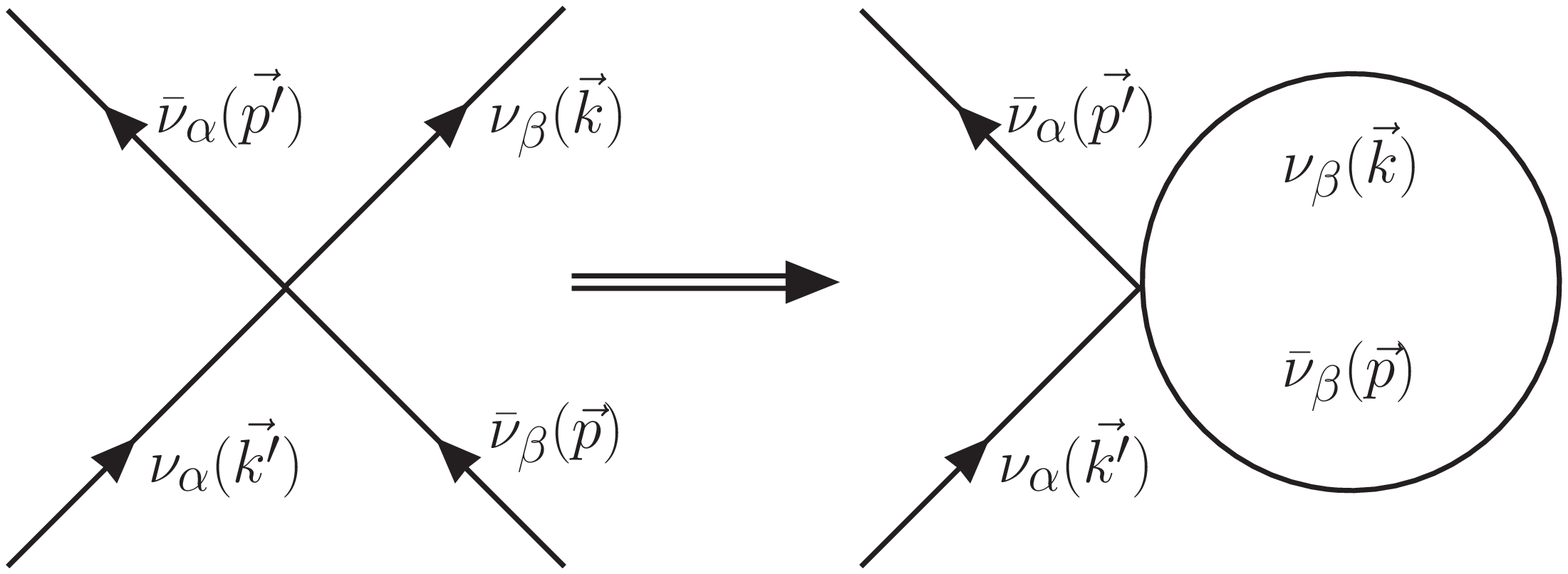}
\includegraphics[scale=0.38]{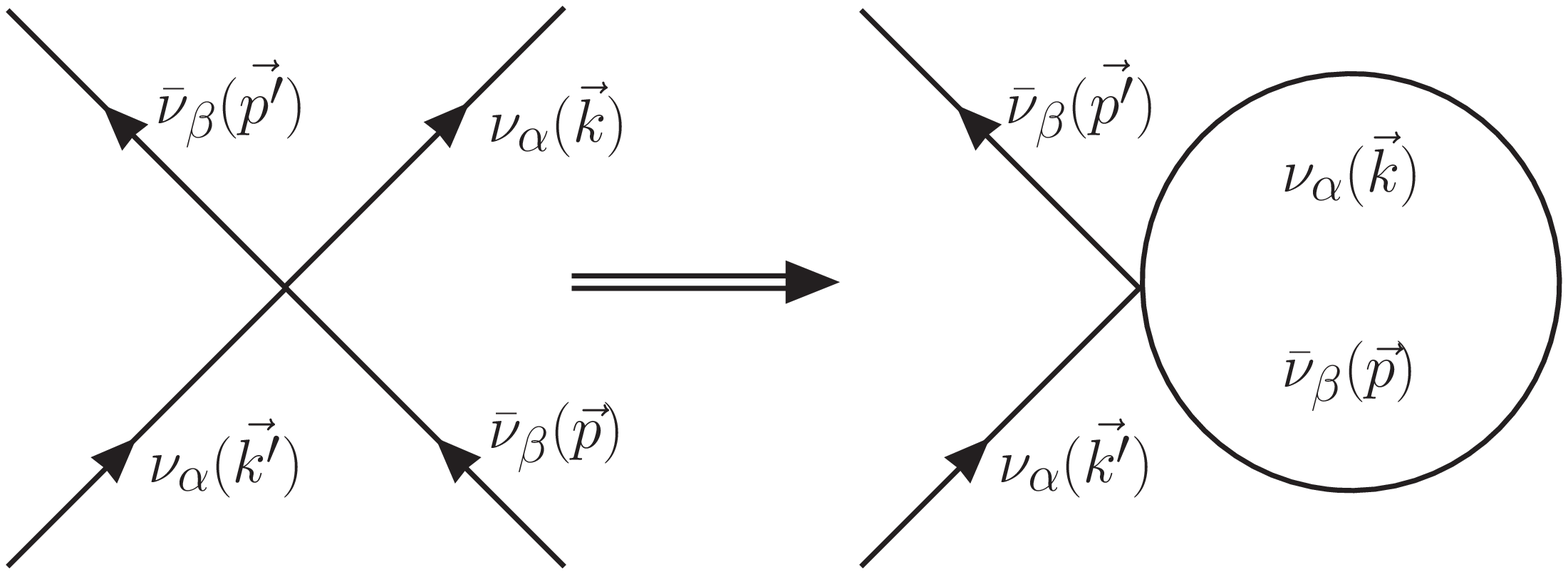}
\end{center}
\caption{The figure shows, in a pictorial way, the interaction of a neutrino with an antineutrino and the corresponding abnormal mean-field $\Delta^*$ Eq.(\ref{e:abpairc}) which is build up from the sum of single-particle neutrino-antineutrino states. Its expression is analogous to the one of the normal mean-field by replacing the normal density by the abnormal one $\kappa$ and by summing over the final (initial in $\Delta$) states. The contributions shown correspond to the diagonal component $\Delta^*_{\nu_{\alpha},\bar{\nu}_{\alpha}}$ Eq.(\ref{e:abpairnusame}) (upper figures), and to the off-diagonal one $\Delta^*_{\nu_{\alpha},\bar{\nu}_{\beta}}$ Eq.(\ref{e:abpairnu}) (lower figures). Note that the diagonal abnormal mean-field also receives a contribution from a term where neutrinos of the same flavor $\alpha$ run in the loop. }
\label{fig:mfnanu}
\end{figure}

We note that  Eqs.(\ref{e:tddm}) can be cast in an elegant and compact matrix form:
\begin{equation}\label{e:50}
i \dot{{\cal R}} = [ {\cal H},{\cal R}],
\end{equation}
where we have introduced the generalized density ${\cal R}$
\begin{equation}\label{e:51}
{\cal R} = \left(
\begin{array}{cc}   
 \rho &  \kappa  \\
\kappa^{\dagger} & 1 - \bar{\rho}^* \end{array}
\right)
\end{equation}
that depends upon both the normal densities for the neutrinos and the antineutrinos and the abnormal density.
The generalized Hamiltonian ${\cal H}$ governing the evolution is given by
\begin{equation}\label{e:52}
{\cal H} = \left(
\begin{array}{cc}   
h & \Delta \\
\Delta^{\dagger} & -\bar{h}^* \end{array}
\right).
\end{equation}
It comprises the mean-field Hamiltonians for the neutrinos and the antineutrinos, as well as the abnormal mean-field.  

\subsection{The abnormal mean-field}
\noindent
Let us now compute the expression of the abnormal mean-field $\Delta$. First, 
the expression for the mean-field $h$ ($\bar{h}^*$) acting on a neutrino (antineutrino) is given by Eqs.(\ref{e:neqmf2}) and (\ref{e:neqmf2anu}). 
To derive an explicit expression for $\Delta$ Eq.(\ref{e:abpair}) and $\Delta^*$ Eq.(\ref{e:abpairc}), one needs to
calculate the matrix element $v_{(ij,kl)}$ associated with the neutral-current interaction of a neutrino with an antineutrino. One obtains the following expression for the off-diagonal components of the abnormal field (see Figure \ref{fig:mfnanu}):
\begin{eqnarray}\label{e:abpairnu}
\Delta^*_{\nu_{\alpha},\bar{\nu}_{\beta}}(\kappa^*_{\nu}) 
 & = & - {{G_F} \over{2\sqrt{2}}}  \int  {{d^3 \vec{p}} \over{(2 \pi)^3  2 E_{p}}}  
\int {{d^3 \vec{k}}\over{(2 \pi)^3  2 E_{k}}} \nonumber \\
&&  (2 \pi)^3 \delta^3(\vec{p} + \vec{k} - \vec{p}\,' -\vec{k}')  \nonumber \\ 
& & [\bar{u}_{\nu_{\alpha}} (\vec{k},h_{\alpha})\gamma_{\mu}(1-\gamma_5) u_{\nu_{\alpha}} (\vec{k}',h'_{\alpha})]  \nonumber \\
& & [\bar{v}_{\nu_{\beta}}  (\vec{p}\,',h'_{\beta})\gamma^{\mu}(1-\gamma_5) v_{\nu_{\beta}} (\vec{p},h_{\beta})]  \nonumber \\ 
& & \langle a^{\dagger}_{\nu_{\alpha}} (\vec{k},h_{\alpha}) b^{\dagger}_{\nu_{\beta} } (\vec{p},h_{\beta}) \rangle
\end{eqnarray}

The explicit expression of the abnormal field Eq.(\ref{e:abpairnu}) depends on the properties of the background:
\beq\label{e:knu}
\kappa^{\nu_{\alpha} \bar{\nu}_{\beta}*}_{\vec{k}h_{\alpha},\vec{p}h_{\beta}} =
\langle a^{\dagger}_{\nu_{\alpha}} (\vec{k},h_{\alpha}) b^{\dagger}_{\nu_{\beta}} (\vec{p},h_{\beta})  \rangle.  
\eeq
Different assumptions can be made on the expectation value of the correlator $a^{\dagger}_{\nu_{\alpha}} (\vec{k},h_{\alpha}) b^{\dagger}_{\nu_{\beta}} (\vec{p},h_{\beta})$
and its complex conjugate. In order to remain as general as possible, various options are considered in the following. 
We give the results for $\Delta^*_{\nu_{\alpha},\bar{\nu}_{\beta}}$ while similar expressions hold for  $\Delta_{\nu_{\alpha},\bar{\nu}_{\beta}}$ as well as
for the diagonal contributions of the abnormal mean-field $\Delta_{\nu_{\alpha},\bar{\nu}_{\alpha}}$ and its complex conjugate $\Delta^*_{\nu_{\alpha},\bar{\nu}_{\alpha}} $ (only the final results are presented here, while more details on their derivation can be found in Appendix C.).
In particular, for the calculation of the latter, 
two contributions need to be added:
\begin{align}\label{e:Ddiagsame}
\Delta^*_{\nu_{\alpha,i},\bar{\nu}_{\alpha,j}}(\kappa^*_{\nu}) & = 2 \Delta^{*(eq)}_{\nu_{\alpha,i},\bar{\nu}_{\alpha,j}}(\kappa_{\nu}^*) + \Delta^{*(uneq)}_{\nu_{\alpha,i},\bar{\nu}_{\alpha,j}}(\kappa^*_{\nu}) \\
&= \sum_{k,l}      [  2  v^{\alpha,\alpha}_{(kl,ij)} \kappa^*_{\nu_{\alpha,k},\bar{\nu}_{\alpha,l}}
+  v^{\alpha,\beta}_{(kl,ij)} \kappa^*_{\nu_{\beta,k},\bar{\nu}_{\beta,l}}]
\end{align}
 as can be seen from Eq.(\ref{e:abpairc}).
The first contribution $\Delta^{*(eq)}_{\nu_{\alpha,i},\bar{\nu}_{\alpha,j}}$ comes from the two amplitudes for the process of neutrino-antineutrino scattering and annihilation for equal flavors (see Figure \ref{fig:mfnanu}) that summed give a factor of 2 times an expression similar to the one shown in 
Eq.(\ref{e:abpairnu}):
\begin{eqnarray}\label{e:abpairnusame}
\Delta^{*(eq)}_{\nu_{\alpha},\bar{\nu}_{\alpha}}(\kappa^*_{\nu}) 
 & = - & {{G_F} \over{2\sqrt{2}}}  \int  {{d^3 \vec{p}} \over{(2 \pi)^3  2 E_{p}}}  
\int {{d^3 \vec{k}}\over{(2 \pi)^3  2 E_{k}}} \nonumber \\
&&
(2 \pi)^3 \delta^3(\vec{p} + \vec{k} - \vec{p}\,' -\vec{k}')   \nonumber \\ 
& & [\bar{u}_{\nu_{\alpha}} (\vec{k},h_{\alpha})\gamma_{\mu}(1-\gamma_5) u_{\nu_{\alpha}} (\vec{k}',h'_{\alpha})]  \nonumber \\
& & [\bar{v}_{\nu_{\alpha}}  (\vec{p}\,',h'_{\alpha})\gamma^{\mu}(1-\gamma_5) v_{\nu_{\alpha}} (\vec{p},h_{\alpha})]  \nonumber \\ 
& & \langle a^{\dagger}_{\nu_{\alpha}} (\vec{p},h_{\alpha}) b^{\dagger}_{\nu_{\alpha}} (\vec{p},h_{\alpha}) \rangle
\end{eqnarray}
The second contribution to $\Delta^{*(uneq)}_{\nu_{\alpha},\bar{\nu}_{\alpha}}$ comes from different flavors than $\alpha$ running in the loop (Fig.\ref{fig:mfnanu}).
Let us now perform an explicit calculation of the pairing mean-field, depending on the background properties.  

The requirement that the background through which neutrinos are traveling is homogeneous, is fulfilled, if we impose that
the neutrino and the antineutrino have  opposite momentum in the expectation values of the neutrino-antineutrino pair operators. 
This corresponds to:
\beq\label{e:knuhompol}
\kappa^{\nu_{\alpha} \bar{\nu}_{\beta}*}_{\vec{k} ,\vec{p} } = (2 \pi)^3 2 E_{k} \delta^3 (\vec{p} + \vec{k})\kappa^{\nu_{\alpha} \bar{\nu}_{\beta}*}_{\vec{p}}
\eeq 
and we take $ h_{\alpha} = -1$ and $h_{\beta} = +1$.
Note that the quantities $\kappa^{\nu_{\alpha} \bar{\nu}_{\beta}*}_{\vec{p}}$ (or $\kappa^{\nu_{\alpha} \bar{\nu}_{\alpha}*}_{\vec{p}}$) are related to the neutrino-antineutrino pair number density: 
\beq\label{e:pair}
\tilde{\kappa}^{\nu_{\alpha} \bar{\nu}_{\beta}*} = \int {d^3\vec{p} \over{(2\pi)^3}} \kappa^{\nu_{\alpha} \bar{\nu}_{\beta}*}_{\vec{p}}
\eeq
in a way analogous e.g. to the electron number densities (see Eq.(\ref{e:enumber})).
From Eq.(\ref{e:abpairnu}), 
by employing the procedure discussed in Section III.A and III.B for the normal fields, one obtains that the abnormal mean-field in spherical coordinates reads
\begin{align}\label{e:Dhompol1}
\Delta^*_{\nu_{\alpha},\bar{\nu}_{\beta}}(\kappa^*_{\nu})  & =  - N(\vec{k}\,',\vec{p}\,')  \sqrt{2} G_F \\ \nonumber
& \quad  \int {d\theta  dp d\phi \over{(2\pi)^3}}  \Bigl[(1+ \cos\theta \cos\theta')\cos(\phi'-\phi) \\ \nonumber
&  \quad  +  i (\cos\theta + \cos\theta')\sin(\phi'-\phi)\\ \nonumber
& \quad +  \sin\theta \sin\theta' \Bigr] \sin\theta p^2 \kappa^{\nu_{\alpha} \bar{\nu}_{\beta}*}_{\vec{p}}
\end{align}
with $N(\vec{k}\,',\vec{p}\,') = (2 \pi)^3 2 E_{k'} \delta^3(\vec{p}\,'+\vec{k}\,')$.
Note that in case there is cylindrical symmetry only the last term remains:
\begin{align}\label{e:Dhompol2}
\Delta_{\nu_{\alpha},\bar{\nu}_{\beta}} (\kappa^*_{\nu}) &= - N(\vec{k}\,',\vec{p}\,')  \sqrt{2} G_F
  \nonumber \\
 & \quad  \int { d\cos\theta dp \over{(2\pi)^2}} \sin\theta \sin\theta' p^2  \kappa^{\nu_{\alpha} \bar{\nu}_{\beta}*}_{\vec{p}}.
\end{align}
The expression Eq.(\ref{e:Dhompol1}) gives in cartesian coordinates is
\begin{align}\label{e:Dhompolcar}
\Delta^*_{\nu_{\alpha},\bar{\nu}_{\beta}} (\kappa^*_{\nu}) & =  - N(\vec{k}\,',\vec{p}\,') \sqrt{2} G_F  \nonumber \\ 
& \quad  \int {d^3\vec{p} \over{(2\pi)^3}}  {{[ d_R(\vec{p},\vec{p}\,') + i d_I(\vec{p},\vec{p}\,')]}\over{|\vec{p}||\vec{p}\, '|\sqrt{p_x^2 + p_y^2}\sqrt{p^{'2}_{x} + p^{'2}_{y}}}}  \kappa^{\nu_{\alpha} \bar{\nu}_{\beta}*}_{\vec{p}}  
\end{align}
with 
\begin{align}\label{e:Dhompolr}
 d_R(\vec{p},\vec{p}\,') =
 & (|\vec{p}||\vec{p}\, '| + p_z p'_{z})(p_x p'_{x} +  p_y p'_{y}) \\ \nonumber
 & + (p_x^2 + p_y^2)(p^{'2}_{x} + p^{'2}_{y})
\end{align}
and
\beq\label{e:Dhompoli}
d_I(\vec{p},\vec{p}\,') =  (p_z|\vec{p}\, '| + p_{z'}|\vec{p}|)(p'_y p_x -p_x p'_{y})
\eeq

So far we have neglected the neutrino masses so that $\nu$ and $\bar{\nu}$ have definite helicities. 
Let us now consider the case that e.g.  $h_{\alpha} = h_{\beta} = +1$ or $ -1$: 
\beq\label{e:knuhomunpol}
\kappa^{\nu_{\alpha} \bar{\nu}_{\beta}*}_{\vec{k} ,\vec{p} } = (2 \pi)^3 2 E_{k} \delta_{h_{\alpha},h_{\beta}} \delta^3 (\vec{p} + \vec{k})\kappa^{\nu_{\alpha} \bar{\nu}_{\beta}*}_{p}
\eeq
While such a contribution is possible, we expect the corresponding abnormal field to be 
suppressed by a factor $(m/E)^2$, $E$ being the neutrino energy. For this specific derivation, since one is sensitive to the neutrino masses, the calculation has to be performed by 
working in the neutrino mass eigenstate basis. 
Following this procedure, one finds
 \begin{align}\label{e:Dhompolcar}
\Delta^*_{\nu_{i'},\bar{\nu}_{j'}}(\kappa^*_{\nu})  & =   (2 \pi)^3 2 \sqrt{E^{i'}_{k'} E^{j'}_{p'}}\delta^3(\vec{p}\,'+\vec{k}\,') {G_F \over{2\sqrt{2}}} \\ \nonumber
& \quad \int {d^3\vec{k} \over{(2\pi)^3}}  \, {{m_{i'} m_i  } \over{E^{i'}_{k'} E^{i}_{k}}} e^{i(\phi'- \phi)}(1 - \cos\theta_{\hat{\vec{k}}\hat{\vec{k}}'}) \kappa^{\nu_{\i} \bar{\nu}_{\j}*}_{\vec{k}}, \\ \nonumber
\end{align}
where here the notation $E^{i'}_{k'}$ indicates the energy of a neutrino with mass $m_{i'}$ and momentum $k'$.

\subsection{The three-flavor neutrino case}
\noindent
We conclude by giving the expressions of the extended dynamical  equations for three neutrino flavors. In this case the abnormal density is
defined as
\beq\label{e:kmat}
\kappa_{\nu} = \left(
\begin{array}{ccc} \langle b_{\nu_e} a_{\nu_e} \rangle  
&  \langle b_{\nu_{\mu}} a_{\nu_e} \rangle & \langle b_{\nu_{\tau}} a_{\nu_e} \rangle \\
\langle b_{\nu_e} a_{\nu_{\mu}} \rangle & \langle b_{\nu_{\mu}} a_{\nu_{\mu}} \rangle
& \langle b_{\nu_{\tau}} a_{\nu_{\mu}} \rangle \\
\langle b_{\nu_e} a_{\nu_{\tau}} \rangle & \langle b_{\nu_{\mu}} a_{\nu_{\tau}} \rangle
& \langle b_{\nu_{\tau}} a_{\nu_{\tau}} \rangle
 \end{array}
\right)
\eeq
and its complex conjugate as
\begin{equation}\label{e:kmat*}
\kappa^*_{\nu}  = \left(
\begin{array}{ccc} 
\langle  a_{\nu_e}^{\dagger} b_{\nu_e}^{\dagger}\rangle  
&  \langle  a_{\nu_e}^{\dagger}b_{\nu_{\mu}}^{\dagger} \rangle & \langle  a_{\nu_e}^{\dagger}b_{\nu_{\tau}}^{\dagger} \rangle \\
\langle  a_{\nu_{\mu}}^{\dagger} b_{\nu_e}^{\dagger}\rangle & \langle  a_{\nu_{\mu}}^{\dagger} b_{\nu_{\mu}}^{\dagger} \rangle
& \langle a_{\nu_{\mu}}^{\dagger} b_{\nu_{\tau}}^{\dagger}\rangle\\
\langle a_{\nu_{\tau}}^{\dagger}b_{\nu_e}^{\dagger}  \rangle & \langle  a_{\nu_{\tau}}^{\dagger} b_{\nu_{\mu}}^{\dagger}\rangle
& \langle  a_{\nu_{\tau}}^{\dagger} b_{\nu_{\tau}}^{\dagger} \rangle
 \end{array}
\right)
\end{equation}
Equation (\ref{e:tddm})  gives to the following set of equations\footnote{Note that here the quantities depend upon only one momentum index.}:
\begin{equation}
\left\{ 
\begin{array}{lcl} 
\vspace{.2cm}
i \dot{\rho}_{\nu_{\alpha}\nu_{\beta}} & =&  [h,\rho]_{\nu_{\alpha}\nu_{\beta}}  \\ 
\vspace{.2cm}
&&  + \sum_i [\Delta_{\nu_{\alpha},\bar{\nu}_{i}}\kappa_{\nu_{\beta},\bar{\nu}_{i}}^* 
- \kappa_{\nu_{\alpha},\bar{\nu}_{i}}\Delta^*_{\nu_{\beta},\bar{\nu}_{i}}] \\
\vspace{.2cm}
i \dot{\bar{\rho}}_{\nu_{\alpha}\nu_{\beta}} & =&  [\bar{h},\bar{\rho}]_{\nu_{\alpha}\nu_{\beta}}  \\ 
\vspace{.2cm}
&& + \sum_i [\Delta_{\nu_{i},\bar{\nu}_{\alpha}}\kappa_{\nu_{i},\bar{\nu}_{\beta}}^* 
- \kappa_{\nu_{i},\bar{\nu}_{\alpha}}\Delta^*_{\nu_{i},\bar{\nu}_{\beta}}] \\
\vspace{.2cm}
i \dot{\kappa}_{\nu_{\alpha}\bar{\nu}_{\beta}} & =& \sum_i h_{\nu_{\alpha}\nu_{i}}\kappa_{\nu_{i}\bar{\nu}_{\beta}} + h_{\bar{\nu}_{\beta}\nu_{i}}
\kappa_{\nu_{\alpha}\nu_{i}} \\
\vspace{.2cm}
&& + \Delta_{\nu_{\alpha},\bar{\nu}_{\beta}} -
 \sum_i [\Delta_{\nu_{i},\bar{\nu}_{\beta}}\rho_{\alpha,\nu_{i}}
+ \Delta_{\nu_{\alpha},\nu_{i}} \bar{\rho}_{\nu_{\beta},\nu_{i}}]\\
\vspace{.2cm}
i \dot{\kappa}^*_{\nu_{\alpha}\bar{\nu}_{\beta}} & =& \sum_i h^*_{\nu_{\alpha}\nu_{i}}\kappa^*_{\nu_{i}\bar{\nu}_{\beta}} + h^*_{\bar{\nu}_{\beta}\nu_{i}}
\kappa^*_{\nu_{\alpha}\nu_{i}} \\
\vspace{.2cm}
&& + \Delta^*_{\nu_{\alpha},\bar{\nu}_{\beta}} -
 \sum_i [\Delta^*_{\nu_{i},\bar{\nu}_{\beta}}\rho^*_{\alpha,\nu_{i}}
+ \Delta^*_{\nu_{\alpha},\nu_{i}} \bar{\rho}^*_{\nu_{\beta},\nu_{i}}]\\
\end{array} \right. 
\end{equation}
where the indices $\alpha,\beta, i = e, \mu, \tau$, $h$ ($\bar{h}$) is given by Eqs.(\ref{e:neqmf}) (Eq.(\ref{e:neqmfanu}))
for neutrinos (antineutrinos), while the $\kappa_{\nu}$ and $\kappa_{\nu}^*$ are given by Eq.(\ref{e:kmat}-\ref{e:kmat*}).

In order to determine the neutrino flavor evolution, one needs to assign initial conditions for the normal and abnormal densities, i.e. 
$\rho_0 = \rho (t=0) $, $\bar{\rho}_0 = \bar{\rho} (t=0) $ and $\kappa_0 = \kappa (t=0)$ for the extended mean-field equations Eq.(\ref{e:50}) with (\ref{e:51}-\ref{e:52}).
In the case of core-collapse supernova neutrinos, such an assignment could be done by extracting the relevant information from realistic simulations of the dense region where neutrinos are trapped that would include the relevant correlations. 
Otherwise, a guess that one can make is to assume that the system is approximately described at initial time by a "stationary state" of the extended mean-field Hamiltonian Eq.(\ref{e:52}); while the system is driven out of such a solution at later times \cite{Volpe:2013}. 

Finally, we would like to point out that, in the application of our equations to the case of supernova neutrinos, the conditions Eqs.(\ref{e:knuhompol}) and (\ref{e:knuhomunpol}) cannot be met, strictly speaking. In fact,  in the region outside the neutrino-sphere, where the density becomes low enough,  all neutrinos and antineutrinos start free streaming. However, in the region just before the neutrino-sphere and at its boundary the presence of collisions might
produce non-zero neutrino-antineutrino correlations as we consider here. It is this intermediate region, that we are interested in, where one goes from the regime  where neutrinos are trapped requiring a Boltzmann treatment, and the one where only forward scattering becomes relevant and the mean-field approximation given by Eqs.(\ref{e:neqmf}-\ref{e:neqmf2anu}) constitutes a good approximation. The presence of $\nu-\bar{\nu}$ correlations
in this transition region might indeed play a role from the point of view of the supernova dynamics, or eventually modify neutrino flavor conversion when neutrinos start free streaming.

\section{Discussion and conclusions}
\noindent
In the present work, we have derived the neutrino evolution equations in a medium composed of matter and eventually of neutrinos, having in mind
environments that produce copious neutrino amounts such as the Sun, core-collapse supernovae, accretion disks around black holes or the early universe. 
In particular, observations of such neutrinos require a precise understanding of the corresponding neutrino number fluxes and spectra, their modifications through such media, and their specific signatures both in solar and supernova neutrino detectors, as well as on stellar, or primordial, nucleosynthesis abundances.  

In the astrophysical environments the neutrino flavor evolution
is essentially treated in the mean-field approximation. While this is a good description in the case of solar neutrinos, the investigation of  neutrino flavor evolution in 
core-collapse supernovae might require going beyond.
For example, in the transition between the dense region where neutrinos are trapped, and the neutrinosphere where they start free streaming, 
many-body correlations might have sizeable effects and impact flavor conversion. The physical argument to support the mean-field assumption is that the MSW resonances occur in the outer layers of the star, because of the large neutrino mass-squared differences, and that the MSW effect is well accounted for in the mean-field approximation. However, this picture -- well separating the region where flavor conversion occurs and the very dense one where the mean-free path characterizes neutrino propagation -- has been modified by recent theoretical studies. The inclusion of the neutrino-neutrino interaction in the treatment of the neutrino propagation, within the mean-field approximation has shown that significant flavor modification can occur in a region close to the neutrinosphere \cite{Duan:2010bg}, with a possible impact on nucleosynthesis \cite{Duan:2010af} and maybe on the explosion dynamics. Ref.\cite{Cherry:2013mv} furnishes an example showing that a careful treatment  of the transition region might be necessary, even from the point of view of the neutrino emission.   
Therefore, although the first investigations show that flavor conversion occurs out of the relevant region to influence the shock waves (see e.g. \cite{Dasgupta:2011jf}), it might still be too early to draw definite conclusions. 

In the context of core-collapse supernovae, a significant step forward beyond the mean-field approximation in the investigation of neutrino flavor conversion would require
the solution of the Boltzmann equation implementing neutrino masses and mixings. Such numerical simulations are demanding, and appear still ahead also within realistic supernova simulations. The situation is different in the context of the early universe, 
since collisions are an essential ingredient bringing neutrino plasma to equilibration, while the mixings bring the system close to flavor equilibration.
The neutrino history is usually determined by solving  
the Boltzmann equation for particles with mixings, or approximate versions of it. 

Several works have addressed many-body aspects of the problem of neutrino
flavor evolution focussing either on the theoretical formulation and the
inherent symmetries \cite{Pehlivan:2011hp}, or on possible implications 
in schematic models (see e.g. \cite{Friedland:2006ke}). 
In the present work we have adopted a novel perspective with respect to previous works and
employed the BBGKY hierarchy as a theoretical framework to go beyond the mean-field approximation, in a realistic treatment of the neutrino interactions with matter and among themselves. In particular we have included
contributions that have not been implemented so far. The BBGKY hierarchy formulation offers a natural truncation scheme with respect to the order of correlations. Such a formulation is equivalent to a description in terms of Green's function in the equal time limit. In particular, the BBGKY hierarchy furnishes a theoretical scheme to go from a many-body treatment to an effective one-body treatment.
We have included, for the first time, novel $\nu\bar{\nu}$ correlations of the pairing type. We have obtained coupled non-linear evolution equations for the normal and abnormal densities, that depend upon the normal and pairing mean-fields.
The explicit expression for such mean-fields has been obtained using the usual
low energy limit for the charged- and neutral-current interactions.  

The abnormal densities involve
the expectation values of the 
bilinear products of the neutrino and antineutrino operators. Such pairs correspond to neutrinos with the same lepton number, while different individual lepton numbers can appear because of the presence of mixings, giving rise to off-diagonal contributions of the abnormal density. Further conditions are imposed to the neutrino-antineutrino pairs,
in the calculation of the abnormal mean-fields, depending on the specific properties of the background that neutrinos are traversing. Requiring that the neutrino-antineutrino pairs have opposite momentum corresponds to a homogeneous medium. For this case we have derived a pairing mean-field with different helicity conditions for the pairs. In particular, 
if one considers contributions from states of positive (negative) helicity for the neutrinos (antineutrinos), 
the pairing mean-field turns out to depend on the ratio, of the neutrino mass  over its energy, squared (as one expects). Although tiny, such off-diagonal contributions might give non-trivial effects. More generally one could consider the general case of an inhomogeneous background. The price to pay  is that one should retain two indices in momentum in the evolution equations and all the normal and abnormal quantities involved. 
 
An hypothesis made in the present work is that the abnormal densities are non-zero at initial time; while
non-zero expectation values for these bilinear operator products
can arise by implementing the collision term in a Boltzmann treatment. Such
a collision term is not accounted for here. However we have been discussing that, in principle, following the BBGKY hierarchy one has the complete evolution equation for the two-body correlation function for our system of neutrinos and antineutrinos including both 
the collision term (also called the Born terms) and the terms dependent on the two-body correlation function (also called the PP terms) that have been assumed to depend upon the abnormal density only in the present work. The three-body correlation function contribution is expected to give higher order corrections. 
Therefore the framework discussed here could also be employed to investigate the relevance of the linked contributions of the two-body correlation function, that
have made the object of the present work, for the case of cosmological neutrinos, at the epoch of Big-bang nucleosynthesis.
 
 Obviously, the procedure we have been describing and employing for our calculations is very general. While it has been used to derive
 dynamical equations for a system of neutrinos propagating in a medium of ordinary matter and neutrinos, applying the procedure to more general cases is straightforward. For example, one can consider that matter is made up of other particles, or that it presents of non-standard interactions. Also, our results can be used for an arbitrary number of neutrino families and, in particular, to account for the presence of sterile neutrinos. 

Finally, in the present manuscript, we have been focussing on the formal aspects implied by 
 the two-body correlations of the neutrino-antineutrino type, and how to implement them in extended evolution equations. Clearly numerical calculations are required to assess the impact of such correlations, or, more generally, of linked contributions of the two-body density matrix (or correlation function)  on neutrino flavor conversion in a medium such as a core-collapse supernova, on nucleosynthesis and maybe the dynamics of these explosive phenomena, or on cosmological neutrinos at the epoch of primordial nucleosynthesis (within a Boltzmann treatment). Further investigations might tell us if these contributions can engender surprising features, or novel mechanisms, in this fascinating domain.

\vspace{.2cm}

\begin{acknowledgments}
\noindent
C. V. thanks J. Serreau for useful discussions and careful reading of this manuscript. C. Espinoza thanks CNRS/IN2P3 for financial support during an early stage of the project, and the Spanish Grant FPA 2008/02878 of Spanish Ministry MICINN. C.E. thanks IPN Orsay for the hospitality during the realization of the project. 

\end{acknowledgments}
\appendix
\section{}
\noindent
For the sake of clarity we give here an explicit formulation of Eq.(\ref{e:wcc12}). Writing the explicit dependence of each quantity on the single-particle configurations,
such an equation reads

\begin{align}
i &\dot{c}_{(ik,jl)} 
\nonumber\\
=&\ [h_1(\rho) + h_2(\rho), c_{12}]_{(ik,jl)} \label{e:c12_1}
\\
\begin{split}
&+ [(1-\rho_1-\rho_2)V(1,2)\rho_{1}\rho_{2}(1-P_{12})]_{(ik,jl)}
\\
&- [(1-P_{12})\rho_{1}\rho_{2}V(1,2)(1-\rho_1-\rho_2)]_{(ik,jl)}
\label{e:c12_2} 
\end{split}
\\
\begin{split}
&+ [(1-\rho_1 -\rho_2)V(1,2) c_{12}]_{(ik,jl)}
\\
&- [c_{12}V(1,2)(1-\rho_1 -\rho_2)]_{(ik,jl)}
\label{e:c12_3} 
\end{split}
\\
&+ \sum_{m=n}[V(1,3),(1-P_{13})\rho_1c_{23}(1-P_{12})]_{(ikm,jln)} 
\label{e:c12_4} 
\\
&+ \sum_{m=n}[V(2,3),(1-P_{23})\rho_2c_{13}(1-P_{12})]_{(ikm,jln)}
\label{e:c12_5}
\end{align}
where the different contributions are :

\begin{align*}
\eqref{e:c12_1} =& \sum_{r}\Bigl[
(t_1 + \Gamma_1)_{ir}c_{(rk,jl)} - c_{(ik,rl)}(t_1 + \Gamma_1)_{rj}
\\
&\qquad+ (t_2 + \Gamma_2)_{kr}c_{(ir,jl)} - c_{(ik,jr)}(t_2 + \Gamma_2)_{rl} 
\Bigr]
\\
\eqref{e:c12_2} =& \sum_{rs}\Bigl\{
v_{(ik,rs)}\rho_{1,rj}\rho_{2,sl}(1-P_{jl})
\\
&\qquad- (1-P_{rs})\rho_{1,ir}\rho_{2,ks}v_{(rs,jl)}
\\
&- \sum_{m=n}\Bigl[
\rho_{1,im}v_{(nk,rs)}\rho_{1,rj}\rho_{2,sl}(1-P_{jl})
\\
&\quad\qquad- (1-P_{rs})\rho_{1,ir}\rho_{2,ks}v_{(rs,ml)}\rho_{1,nj}
\\
&\quad\qquad+ \rho_{2,km}v_{(in,rs)}\rho_{1,rj}\rho_{2,sl}(1-P_{jl})
\\
&\quad\qquad- (1-P_{rs})\rho_{1,ir}\rho_{2,ks}v_{(rs,jm)}\rho_{2,nl}
\Bigr]
\Bigr\}
\\
\eqref{e:c12_3} =& \sum_{rs}\Bigl\{
v_{(ik,rs)} c_{(rs,jl)} - c_{(ik,rs)} v_{(rs,jl)}
\\
&-\sum_{m=n} \Bigl[
\rho_{1,im} v_{(nk,rs)} c_{(rs,jl)}
- c_{(ik,rs)}v_{(rs,ml)}\rho_{1,nj}
\\
&\quad\qquad+ \rho_{2,km} v_{(in,rs)} c_{(rs,jl)}
- c_{(ik,rs)}v_{(rs,jm)}\rho_{2,nl}
\Bigr]
\Bigr\}
\\
\eqref{e:c12_4} =& \sum_{rs}\sum_{m=n} \Bigl[
v_{(im,rs)}(1-P_{rs})\rho_{1,rj}c_{(ks,ln)}(1-P_{jl})
\\
&\qquad\qquad- (1-P_{rs})\rho_{1,ir} c_{(km,ls)}(1-P_{ik})v_{(rs,jn)}
\Bigr]
\\
\eqref{e:c12_5} =& \sum_{rs}\sum_{m=n} \Bigl[
v_{(km,rs)}(1-P_{rs})\rho_{2,rl}c_{(is,jn)}(1-P_{jl})
\\
&\qquad\qquad- (1-P_{rs})\rho_{2,kr}c_{(im,js)}(1-P_{ik})v_{(rs,ln)}
\Bigr]
\end{align*}
\section{Derivation of the normal mean-field}
\noindent
For completeness, we write down intermediate results in the calculation of the normal fields for the case of neutrinos interacting with antineutrinos, while the same procedure
leads to all the expressions of the normal mean-fields given in Section III.
Let us consider the case of the off-diagonal contribution of $\Gamma$ Eq.(\ref{e:nnu}), but associated with neutrino of flavor $\nu_{\alpha}$ traversing a medium of 
antineutrinos  of a different flavor $\nu_{\beta}$ (Fig.\ref{fig:mfvv}) :
\begin{eqnarray}\label{e:nnu2}
\Gamma_{\nu_{\alpha},\nu_{\beta}}(\bar{\rho}_{\nu}) 
 & = & - \int  {{d^3 \vec{p}} \over{(2 \pi)^3  2 E_{p}}}  
\int {{d^3 \vec{p}\,'}\over{(2 \pi)^3  2 E_{p'}}} \nonumber \\
&&{{G_F} \over{2\sqrt{2}}} 
\int d^3 \vec{x} ~ e^{i(\vec{p} + \vec{k} - \vec{p}' -\vec{k'}) \cdot \vec{x}} \nonumber \\ 
& & [\bar{u}_{\nu_{\alpha}} (\vec{k},h_{\alpha})\gamma_{\mu}(1-\gamma_5) u_{\nu_{\beta}} (\vec{k}',h'_{\beta})]  \nonumber \\
& & [\bar{v}_{\nu_{\beta}}  (\vec{p}\,',h'_{\beta})\gamma^{\mu}(1-\gamma_5) v_{\nu_{\alpha}} (\vec{p},h_{\alpha})]  \nonumber \\ 
& & \langle  b^{\dagger}_{\nu_{\alpha}} (\vec{p},h) b_{\nu_{\beta} } (\vec{p}\,',h')\rangle
\end{eqnarray}
Note the minus sign that comes from the fact that there are antineutrinos in the background.
Implementing that the expectation value over the background :
\beq\label{e:rhobar}
\bar{\rho}^{\nu_{\beta}, \nu_{\alpha}}_{\vec{p}\,'h',\vec{p}h} \equiv  
\langle  b^{\dagger}_{\nu_{\alpha}} (\vec{p},h) b_{\nu_{\beta}} (\vec{p}\,',h')\rangle
\eeq
satisfies :
\beq\label{e:rhoanuh}
\bar{\rho}^{\nu_{\beta}, \nu_{\alpha}}_{\vec{p}\,'h',\vec{p}h}= (2 \pi)^3 2 E_{p} \delta_{hh'} \delta^3 (\vec{p} -\vec{p}\,')\bar{\rho}^{\nu_{\beta}, \nu_{\alpha}}_{p}
\eeq
one gets :
\begin{align}
\Gamma_{\nu_{\alpha},\nu_{\beta}}(\bar{\rho}_{\nu})  
&= - N(\vec{k},\vec{k'}) \ {{G_F} \over{2\sqrt{2}}}  \int {{d^3 \vec{p}}\over{(2 \pi)^3  2 E_{p}}}\ \bar{\rho}^{\nu_{\beta}, \nu_{\alpha}}_{p}
\nonumber \\ 
&\quad [\bar{u}_{\nu_{\alpha}} (\vec{k},-)\gamma_{\mu}(1-\gamma_5) u_{\nu_{\alpha'}} (\vec{k},-)]  
\nonumber \\
&\quad [\bar{v}_{\bar{\nu}_{\beta'}}  (\vec{p},+)\gamma^{\mu}(1-\gamma_5) v_{\bar{\nu}_{\beta}} (\vec{p},+)]
\end{align}
with $N(\vec{k},\vec{k'}) = (2 \pi)^3\,\delta^3(\vec{k}-\vec{k'})$.
By using the well known trace relations :
\begin{eqnarray}
&&\bar{u}_{\nu_{\alpha}} (\vec{k},h)\gamma_{\eta}(1-\gamma_5) u_{\nu_{\alpha'}} (\vec{k},h) 
\nonumber\\ 
&&=
\mathrm{Tr}\left[
u_{\nu_{\alpha'}} (\vec{k},h)\bar{u}_{\nu_{\alpha}} (\vec{k},h)\gamma_{\eta}(1-\gamma_5)  
\right]
\nonumber\\ 
&&=
\mathrm{Tr}\left[
(\slashed{k} + m_{\nu})\left(\frac{1+\gamma^5\slashed{s}(k)}{2}\right)\gamma_{\eta}(1 - \gamma^5)
\right]
\nonumber\\ 
&&=
2 \left( k_{\eta} - m_{\nu}\, s_{\eta}({k}) \right)
\end{eqnarray} 
with
\[
s_{\eta}(k) = \frac{h}{m_{\nu}}\left({|\vec{k}|},E_{k}\frac{\vec{k}}{|\vec{k}|}\right)
\]
the expression becomes\footnote{Note that we show terms with (effective) masses while we have considered relativistic neutrinos and neglected such contributions in the calculations.}: 
\begin{align}
\Gamma_{\nu_{\alpha},\nu_{\beta}}(\bar{\rho}_{\nu})  
&= - N(\vec{k},\vec{k'}) \ 
{{G_F} \over{2\sqrt{2}}}  \int {{d^3 \vec{p}}\over{(2 \pi)^3  2 E_{p}}}
\nonumber\\
&\quad 4
\left[ k_{\eta} - m_{\nu}\,s_{\eta}({k}) \right]
\left[ p^{\eta} + m_{\bar{\nu}}\,s^{\eta}({p}) \right]
\ \bar{\rho}^{\nu_{\beta}, \nu_{\alpha}}_{p}
\nonumber \\ 
&= - N(\vec{k},\vec{k'}) \
{{G_F} \over{2\sqrt{2}}}  \int {{d^3 \vec{p}}\over{(2 \pi)^3  2 E_{p}}}
\nonumber \\ 
&\quad
4\Bigl[k\cdot p - m_{\nu}m_{\bar{\nu}}\,s({k})\cdot s({p}) \nonumber\\
&\qquad
- m_{\nu}\, s({k})\cdot p + m_{\bar{\nu}}\, k\cdot s({p}) 
\Bigr]\ \bar{\rho}^{\nu_{\beta}, \nu_{\alpha}}_{p} \ ,
\end{align}
which finally gives for relativistic neutrinos the known relation :
\begin{align}
\Gamma_{\nu_{\alpha},\nu_{\beta}}(\bar{\rho}_{\nu})  &= - 2 E_{k'}N(\vec{k},\vec{k'}) \ \sqrt{2}{G_F}\int {{d^3 \vec{p}}\over{(2 \pi)^3}}\ (1 - \hat{k}\cdot\hat{p}) \bar{\rho}^{\nu_{\alpha}, \nu_{\beta} *}_{p}
\end{align}
with $\bar{\rho}^{\nu_{\alpha}, \nu_{\beta} *}_{p} = \bar{\rho}^{\nu_{\beta}, \nu_{\alpha}}_{p}$.

\section{Derivation of the abnormal mean-field}
\noindent
For the sake of clarity we give here more intermediate steps in the calculation of the abnormal mean-fields. Let us first consider the homogeneous case with neutrinos (antineutrinos) described only by negative (positive) helicity eigenstates. The off-diagonal contribution to the pairing potential is
\begin{align}
&\Delta_{\nu_{\alpha}(\vec{k'},-),\bar{\nu}_{\beta}(\vec{p}\,',+)}^* 
\nonumber\\
&\quad= 
 -  \int  {{d^3 \vec{k}} \over{(2 \pi)^3  2 E_{k}}}  
\int {{d^3 \vec{p}}\over{(2 \pi)^3  2 E_{p}}} 
\nonumber \\
&\qquad  {{G_F} \over{2\sqrt{2}}} \int d^3 \vec{x} ~ e^{i(\vec{p} + \vec{k} - \vec{p}' -\vec{k'}) \cdot \vec{x}} 
\nonumber \\ 
&\qquad [\bar{u}_{\nu_{\alpha}} (\vec{k},-)\gamma_{\mu}(1-\gamma_5) u_{\nu_{\alpha}} (\vec{k}',-)]  \nonumber \\
&\qquad [\bar{v}_{\nu_{\beta}}  (\vec{p}\,',+)\gamma^{\mu}(1-\gamma_5) v_{\nu_{\beta}} (\vec{p}\, ,+)]  \nonumber \\ 
&\qquad \langle a^{\dagger}_{\nu_{\alpha}} (\vec{k},-) b^{\dagger}_{\nu_{\beta}} (\vec{p},+) \rangle
\end{align}
Implementing the homogeneity condition Eq.~\eqref{e:knuhompol} and performing a Fierz transformation result in
\begin{align}
&\Delta_{\nu_{\alpha}(\vec{k'},-),\bar{\nu}_{\beta}(-\vec{k'},+)}^* 
\nonumber\\
&\quad= +\ (2 \pi)^3\,\delta^3(\vec{p}\,'+\vec{k'})\ {{G_F} \over{2\sqrt{2}}}  
\int {{d^3 \vec{k}}\over{(2 \pi)^3  2 E_{k}}}
\nonumber \\ 
&\qquad [\bar{u}_{\nu_{\alpha}} (\vec{k},-)\gamma_{\mu}(1-\gamma_5) v_{\bar{\nu}_{\beta}} (-\vec{k},+)]  
\nonumber \\
&\qquad [\bar{v}_{\bar{\nu}_{\beta}}  (-\vec{k'},+)\gamma^{\mu}(1-\gamma_5) u_{\nu_{\alpha}} (\vec{k'},-)]\ \kappa^{\nu_{\alpha}\bar{\nu}_{\beta}*}_{\vec{k}}
\label{e:adelta1}
\end{align}

The above expression can be evaluated using the following definitions. We use the chiral representation.
The four-component spinors can be expressed as
\begin{align}
u(k,h) =& 
\begin{pmatrix}
-\sqrt{E+h|\vec{k}|}\ \chi_h(\vec{k}) \\
\sqrt{E-h|\vec{k}|}\ \chi_h(\vec{k})
\end{pmatrix} \ ,
\\
v(k,h) =& -h
\begin{pmatrix}
\sqrt{E-h|\vec{k}|}\ \chi_{-h}(\vec{k}) \\
\sqrt{E+h|\vec{k}|}\ \chi_{-h}(\vec{k})
\end{pmatrix} \ .
\end{align}
with the two-component helicity eigenstate spinors given in spherical coordinates 
(polar angle $\theta$ and azimuthal angle $\phi$) by
\begin{align}
\chi_{+}(\vec{k}) &=
\begin{pmatrix}
\cos\frac{\theta}{2} \\
\sin\frac{\theta}{2}\,e^{i\phi}
\end{pmatrix} \ ,
\quad
\chi_{-}(\vec{k}) &=
\begin{pmatrix}
-\sin\frac{\theta}{2}\,e^{-i\phi} \\
\cos\frac{\theta}{2}
\end{pmatrix} \ ,
\end{align}
which satisfy the useful relations:
\begin{align}
\chi_{+}(-\vec{k}) &= -e^{i\phi} \chi_{-}(\vec{k})
\ , \quad
\chi_{-}(-\vec{k}) = e^{-i\phi} \chi_{+}(\vec{k}) \ .
\end{align}

\noindent
By using the Dirac gamma matrices 
\begin{align}
\gamma^0 &=
\begin{pmatrix}
0 & - {\bf 1} \\
- {\bf 1} & 0
\end{pmatrix} \ ,
\
\vec{\gamma} = 
\begin{pmatrix}
0 & \vec{\bf \sigma} \\
-\vec{\bf \sigma} & 0
\end{pmatrix} \ ,
\
\gamma^5 = 
\begin{pmatrix}
{\bf 1} & 0 \\
0 & -{\bf 1}
\end{pmatrix} 
\end{align}
and the Pauli spin matrices :
\begin{equation}
\vec{\bf \sigma} =
\left(
\begin{pmatrix}
0 & 1 \\
1 & 0
\end{pmatrix} ,
i
\begin{pmatrix}
0 & -1 \\
1 & 0
\end{pmatrix} ,
\begin{pmatrix}
1 & 0 \\
0 & -1
\end{pmatrix}
\right) \ .
\end{equation}
one obtains for the space-component  of the first spinor product in Eq.~\eqref{e:adelta1}:
\begin{align}
&\bar{u}_{\nu_{\alpha}} (\vec{k},-)\vec{\gamma}(1-\gamma_5) v_{\bar{\nu}_{\beta}} (-\vec{k},+) 
\nonumber\\ 
&= 4E\, \chi^{\dagger}_{-}(\vec{k})\, \vec{\bf \sigma}\, \chi_{-}(-\vec{k})
\nonumber\\
&= 4E\,  
\Bigl( 
\cos\theta\cos\phi - i\sin\phi , 
\cos\theta\sin\phi + i\cos\phi , 
-\sin\theta
 \Bigr)
\\
&= \frac{4E}{|\vec{k}|\sqrt{k_x^2 + k_y^2}}\,
\Bigl( 
k_x k_z - i k_y|\vec{k}|\ , k_z k_y + i k_x|\vec{k}|\ ,
-(k_x^2 + k_y^2)
 \Bigr) \ ,
\end{align} 
while the time-component $(\gamma_{\mu} = \gamma_{0})$ vanishes.
The second spinor product $\bar{v}_{\bar{\nu}_{\beta}}  (-\vec{k'},+)\gamma^{\mu}(1-\gamma_5) u_{\nu_{\alpha}} (\vec{k'},-)$ gives simply a complex conjugate of the above result. Substituting these expressions into the Eq.~\eqref{e:adelta1} one obtains the results Eqs.~\eqref{e:Dhompol1} and~\eqref{e:Dhompolcar}. 

Since for the calculation of the abnormal field Eq.(\ref{e:Dhompolcar}) we also use the following type spinor products:
\beq
S1 = \bar{u}_{\nu_{a}} (\vec{k},+)\gamma_{\mu}(1-\gamma_5) v_{\bar{\nu}_{b}} (-\vec{k},+) \ ,
\eeq
and
\beq
S2 = \bar{u}_{\nu_{a}} (\vec{k},-)\gamma_{\mu}(1-\gamma_5) v_{\bar{\nu}_{b}} (-\vec{k},-) \ ,
\eeq
where $a$ and $b$ refer to given neutrino mass eigenstate, 
we give the resuls for the time- and the space-components:
\begin{align}
&\bar{u}_{\nu_{a}} (\vec{k},+)\gamma_{0}(1-\gamma_5) v_{\bar{\nu}_{b}} (-\vec{k},+)
\nonumber\\
&= - 2m_{a}\sqrt{\frac{E_b}{E_a}}e^{-i\phi}
\\
&= - 2m_{a}\sqrt{\frac{E_{b}}{E_{a}}}\,\frac{k_x - i k_y}{\sqrt{k_x^2 + k_y^2}}
\end{align}
\begin{align}
&\bar{u}_{\nu_{a}} (\vec{k},+)\vec{\gamma}(1-\gamma_5) v_{\bar{\nu}_{b}} (-\vec{k},+)
\nonumber\\
&= 2 m_a \sqrt{\frac{E_d}{E_b}}e^{-i\phi}(\sin\theta\cos\phi,\sin\theta\sin\phi,\cos\theta)
\\
&= 2 m_a \sqrt{\frac{E_d}{E_b}}\frac{k_x - i k_y}{\sqrt{k_x^2 + k_y^2}}\frac{\vec{k}}{|\vec{k}|}
\end{align}
\noindent
Similarly for the time and space components of $S2$ one obtains: 
\begin{align}
&\bar{u}_{\nu_{a}} (\vec{k},-)\gamma_{0}(1-\gamma_5) v_{\bar{\nu}_{b}} (-\vec{k},-)
\nonumber\\
&= -2m_{b}\sqrt{\frac{E_a}{E_b}}e^{+i\phi}
\\
&= -2m_{b}\sqrt{\frac{E_{a}}{E_{b}}}\,\frac{k_x + i k_y}{\sqrt{k_x^2 + k_y^2}}
\end{align}
\begin{align}
&\bar{u}_{\nu_{a}} (\vec{k},-)\vec{\gamma}(1-\gamma_5) v_{\bar{\nu}_{b}} (-\vec{k},-)
\nonumber\\
&= -2m_b\sqrt{\frac{E_a}{E_b}}e^{+i\phi}(\sin\theta\cos\phi,\sin\theta\sin\phi,\cos\theta)
\\
&= -2m_{b}\sqrt{\frac{E_{a}}{E_{b}}}\,\frac{k_x + i k_y}{\sqrt{k_x^2 + k_y^2}}\frac{\vec{k}}{|\vec{k}|}
\end{align}
In the last expressions we have employed the following relativistic expressions for the spinors:
\begin{align}
u(k,-) &\approx \sqrt{2E}
\begin{pmatrix} 
-\frac{m}{2E}\ \chi_{-}(\vec{k}) \\
\ \chi_{-}(\vec{k})
\end{pmatrix} \ ,
\\
u(k,+) &\approx -\sqrt{2E}
\begin{pmatrix} 
\chi_{+}(\vec{k}) \\
-\frac{m}{2E}\ \chi_{+}(\vec{k})
\end{pmatrix} \ ,
\\
v(k,+) &\approx -
\sqrt{2E}
\begin{pmatrix} 
\frac{m}{2E}\ \chi_{-}(\vec{k}) \\
\chi_{-}(\vec{k})
\end{pmatrix} \ ,
\\
v(k,-) &\approx \sqrt{2E}
\begin{pmatrix} 
\chi_{+}(\vec{k}) \\
\frac{m}{2E}\ \chi_{+}(\vec{k})
\end{pmatrix} \ .
\end{align}




\begin{thebibliography}{99}
\bibitem{Pontecorvo:1957cp} 
  B.~Pontecorvo,
  J. Exptl. Theoret. Phys. {\bf 33}, 549 (1957)
  [ Sov. Phys. JETP {\bf 6}, 429 (1958)].


\bibitem{Fukuda:1998ah} 
  Y.~Fukuda {\it et al.}  [Super-Kamiokande Collaboration],
  Phys.\ Rev.\ Lett.\  {\bf 82}, 2644 (1999)
  [hep-ex/9812014].

\bibitem{PDG2012} 
J. Beringer et al. [Particle Data Group], Phys.\ Rev.\ D {\bf 86}, 
010001 (2012).

\bibitem{Maki:1962mu} 
  Z.~Maki, M.~Nakagawa and S.~Sakata,
  Prog.\ Theor.\ Phys.\  {\bf 28}, 870 (1962).


\bibitem{Giunti:2007ry}
  C.~Giunti and C.~W.~Kim,
   ``Fundamentals of Neutrino Physics and Astrophysics,''
Oxford University Press 2007.

\bibitem{Wolfenstein1977} L.~Wolfenstein, Phys. Rev., {\bf D17}, 2369 (1978).

\bibitem{M&S1986} S.~P.~Mikheev and A.~I.~Smirnov, Nuovo Cimento, {\bf 9C}, 17, (1986).


\bibitem{Pantaleone:1992eq}
  J.~T.~Pantaleone,
  Phys.\ Lett.\  B, {\bf 287}, 128 (1992).

\bibitem{Samuel:1993uw}
  S.~Samuel,
  \prd, {\bf 48}, 1462 (1993).


\bibitem{Pastor:2001iu}
  S.~Pastor, G.~G.~Raffelt and D.~V.~Semikoz,
  Phys.\ Rev.\  D {\bf 65}, 053011 (2002)
  [arXiv:hep-ph/0109035].

\bibitem{Duan:2007mv} 
  H.~Duan, G.~M.~Fuller, J.~Carlson and Y.~-Z.~Qian,
  Phys.\ Rev.\ D {\bf 75}, 125005 (2007)
  [astro-ph/0703776].


\bibitem{Hannestad:2006nj}
  S.~Hannestad, G.~G.~Raffelt, G.~Sigl and Y.~Y.~Y.~Wong,
  Phys.\ Rev.\  D {\bf 74}, 105010 (2006)
  [Erratum-ibid.\  D {\bf 76}, 029901 (2007)]
  [arXiv:astro-ph/0608695].

\bibitem{Galais:2011gh} 
  S.~Galais and C.~Volpe,
  Phys.\ Rev.\ D {\bf 84}, 085005 (2011)
  [arXiv:1103.5302 [astro-ph.SR]].


\bibitem{Duan:2010bg} 
  H.~Duan, G.~M.~Fuller and Y.~-Z.~Qian,
  Ann.\ Rev.\ Nucl.\ Part.\ Sci.\  {\bf 60}, 569 (2010)
  [arXiv:1001.2799 [hep-ph]].

\bibitem{Dasgupta:2005wn} 
  B.~Dasgupta and A.~Dighe,
  Phys.\ Rev.\ D {\bf 75}, 093002 (2007)
  [hep-ph/0510219].

\bibitem{Fogli:2006xy} 
  G.~L.~Fogli, E.~Lisi, A.~Mirizzi and D.~Montanino,
  JCAP {\bf 0606}, 012 (2006)
  [hep-ph/0603033].


\bibitem{Kneller:2010sc} 
  J.~P.~Kneller and C.~Volpe,
  Phys.\ Rev.\ D {\bf 82}, 123004 (2010)
  [arXiv:1006.0913 [hep-ph]].



\bibitem{Dolgov:2002wy} 
  A.~D.~Dolgov,
  Phys.\ Rept.\  {\bf 370}, 333 (2002)
  [hep-ph/0202122].

\bibitem{Iocco:2008va} 
  F.~Iocco, G.~Mangano, G.~Miele, O.~Pisanti and P.~D.~Serpico,
  Phys.\ Rept.\  {\bf 472}, 1 (2009)
  [arXiv:0809.0631 [astro-ph]].

\bibitem{Mangano:2010ei} 
  G.~Mangano, G.~Miele, S.~Pastor, O.~Pisanti and S.~Sarikas,
  JCAP {\bf 1103}, 035 (2011)
  [arXiv:1011.0916 [astro-ph.CO]].

\bibitem{Dolgov:2002ab} 
  A.~D.~Dolgov, S.~H.~Hansen, S.~Pastor, S.~T.~Petcov, G.~G.~Raffelt and D.~V.~Semikoz,
  Nucl.\ Phys.\ B {\bf 632}, 363 (2002)
  [hep-ph/0201287].

\bibitem{Abazajian:2002qx} 
  K.~N.~Abazajian, J.~F.~Beacom and N.~F.~Bell,
  Phys.\ Rev.\ D {\bf 66}, 013008 (2002)
  [astro-ph/0203442].

\bibitem{Gava:2010kz} 
  J.~Gava and C.~Volpe,
  Nucl.\ Phys.\ B {\bf 837}, 50 (2010)
  [arXiv:1002.0981 [hep-ph]].

\bibitem{Mirizzi:2012we} 
  A.~Mirizzi, N.~Saviano, G.~Miele and P.~D.~Serpico,
  Phys.\ Rev.\ D {\bf 86}, 053009 (2012)
  [arXiv:1206.1046 [hep-ph]].

\bibitem{Dolgov:2003sg} 
  A.~D.~Dolgov and F.~L.~Villante,
  Nucl.\ Phys.\ B {\bf 679}, 261 (2004)
  [hep-ph/0308083].

\bibitem{Abazajian:2004aj} 
  K.~Abazajian, N.~F.~Bell, G.~M.~Fuller and Y.~Y.~Y.~Wong,
  Phys.\ Rev.\ D {\bf 72}, 063004 (2005)
  [astro-ph/0410175].

\bibitem{Hannestad:2012ky} 
  S.~Hannestad, I.~Tamborra and T.~Tram,
  JCAP {\bf 1207}, 025 (2012)
  [arXiv:1204.5861 [astro-ph.CO]].


\bibitem{Dolgov:1980cq} 
  A.~D.~Dolgov,
  Sov.\ J.\ Nucl.\ Phys.\  {\bf 33}, 700 (1981)
  [Yad.\ Fiz.\  {\bf 33}, 1309 (1981)].

\bibitem{Raffelt:1992uj} 
  G.~Raffelt, G.~Sigl and L.~Stodolsky,
  Phys.\ Rev.\ Lett.\  {\bf 70}, 2363 (1993)
  [Erratum-ibid.\  {\bf 98}, 069902 (2007)]
  [hep-ph/9209276].

 
\bibitem{Sigl:1992fn} 
  G.~Sigl and G.~Raffelt,
  Nucl.\ Phys.\ B {\bf 406}, 423 (1993).

\cite{Qian:1994wh}
\bibitem{Qian:1994wh} 
  Y.~Z.~Qian and G.~M.~Fuller,
  Phys.\ Rev.\ D {\bf 51}, 1479 (1995)
  [astro-ph/9406073].
 
\bibitem{McKellar:1992ja} 
  B.~H.~J.~McKellar and M.~J.~Thomson,
  Phys.\ Rev.\ D {\bf 49}, 2710 (1994).

 
\bibitem{Sawyer:2005jk} 
  R.~F.~Sawyer,
  Phys.\ Rev.\ D {\bf 72}, 045003 (2005)
  [hep-ph/0503013].

  
 

\bibitem{Friedland:2003dv} 
  A.~Friedland and C.~Lunardini,
  Phys.\ Rev.\ D {\bf 68}, 013007 (2003)
  [hep-ph/0304055].

\bibitem{Friedland:2003eh} 
  A.~Friedland and C.~Lunardini,
  JHEP {\bf 0310}, 043 (2003)
  [hep-ph/0307140].

\bibitem{Friedland:2006ke} 
  A.~Friedland, B.~H.~J.~McKellar and I.~Okuniewicz,
  Phys.\ Rev.\ D {\bf 73}, 093002 (2006)
  [hep-ph/0602016].


\bibitem{Cardall:2007zw} 
  C.~Y.~Cardall,
  Phys.\ Rev.\ D {\bf 78}, 085017 (2008)
  [arXiv:0712.1188 [astro-ph]].



\bibitem{Balantekin:2006tg} 
  A.~B.~Balantekin and Y.~Pehlivan,
  J.\ Phys.\ G {\bf 34}, 47 (2007)
  [astro-ph/0607527].

\bibitem{Pehlivan:2011hp} 
  Y.~Pehlivan, A.~B.~Balantekin, T.~Kajino and T.~Yoshida,
  Phys.\ Rev.\ D {\bf 84}, 065008 (2011)
  [arXiv:1105.1182].

\bibitem{Born-Green}
M. Born and H.S. Green, Proc. \ Roy. \ Soc.\ A {\bf 188},10 (1946).

\bibitem{Yvon}
J. \ Yvon, Act. \ Sci.\ et Ind.\, 203 (1935).

\bibitem{Kirkwood}
Kirkwood, J. \ Chem. \ Phys. \ 3, 300 (1935).

\bibitem{Bogoliubov}
N.N. Bogoliubov, Journal of Physics USSR, 10, 265 (1946).




\bibitem{Simenel:2008mh} 
  C.~Simenel, B.~Avez and D.~Lacroix,
  arXiv:0806.2714 [nucl-th].


\bibitem{Ring} 
  P. Ring and P. Schuck,
  ``The nuclear many body problem",
  Springer Verlag Edition (1990).

\bibitem{wang85} 
S.-J. Wang and W. Cassing,
Annals of physics {\bf 159}, 328 (1985).

\bibitem{lacroix} D. \ Lacroix, S. Ayik, P. \ Chomaz,
Progress in Part. and Nucl. Physics 52 (2004) 497.



\bibitem{Wang:1994xc} 
  S.~-J.~Wang, W.~Zuo and W.~Cassing,
  Nucl.\ Phys.\ A {\bf 573}, 245 (1994)
  [nucl-th/9401012].

\bibitem{Liebendoerfer:2003es}
 M.~Liebendoerfer, M.~Rampp, H.~-T.~.Janka and A.~Mezzacappa,
of methods,''
 Astrophys.\ J.\  {\bf 620}, 840 (2005)
 [astro-ph/0310662].

\bibitem{Mezzacappa:2005ju}
 A.~Mezzacappa,
 Ann.\ Rev.\ Nucl.\ Part.\ Sci.\  {\bf 55}, 467 (2005).

\bibitem{Woosley:2005}
 S.~Woosley, H-T.~Janka,
 Nature \ Physics \ {\bf 1}, 147 - 154 (2005).

\bibitem{Kotake:2005zn}
 K.~Kotake, K.~Sato and K.~Takahashi,
core-collapse supernovae,''
 Rept.\ Prog.\ Phys.\  {\bf 69}, 971 (2006)
 [astro-ph/0509456].

\bibitem{Janka:2012}
 H-T. Janka,
 Ann. \ Rev. \ Nucl. \ Part. \ Sci. {\bf 62}, 407-451 (2012). 

\bibitem{Giunti:1991cb} 
  C.~Giunti, C.~W.~Kim and U.~W.~Lee,
  Phys.\ Rev.\ D {\bf 45}, 2414 (1992).

\bibitem{Kostelecky:1993yt} 
  V.~A.~Kostelecky, J.~T.~Pantaleone and S.~Samuel,
  Phys.\ Lett.\ B {\bf 315}, 46 (1993).

\bibitem{Fidler:2011yq} 
  C.~Fidler, M.~Herranen, K.~Kainulainen and P.~M.~Rahkila,
  JHEP {\bf 1202}, 065 (2012)
  [arXiv:1108.2309 [hep-ph]].

\bibitem{deGouvea:2012hg} 
  A.~de Gouvea and S.~Shalgar,
  JCAP {\bf 1210}, 027 (2012)
  [arXiv:1207.0516 [astro-ph.HE]].

\bibitem{Sawyer:2010jk} 
  R.~F.~Sawyer,
  Phys.\ Rev.\ D {\bf 83}, 065023 (2011)
  [arXiv:1011.4585 [astro-ph.CO]].


\bibitem{Volpe:2013} 
D.V\"a\"an\"anen and C. Volpe, work in progress.
  


\bibitem{Cherry:2013mv} 
  J.~F.~Cherry, J.~Carlson, A.~Friedland, G.~M.~Fuller and A.~Vlasenko,
  arXiv:1302.1159 [astro-ph.HE].

\bibitem{Duan:2010af} 
  H.~Duan, A.~Friedland, G.~C.~McLaughlin and R.~Surman,
  J.\ Phys.\ G {\bf 38}, 035201 (2011)
  [arXiv:1012.0532 [astro-ph.SR]].


\bibitem{Dasgupta:2011jf} 
  B.~Dasgupta, E.~P.~O'Connor and C.~D.~Ott,
  Phys.\ Rev.\ D {\bf 85}, 065008 (2012)
  [arXiv:1106.1167 [astro-ph.SR]].
  
 
\end{thebibliography}
\end{document}